\documentclass[prb,twocolumn,showpacs,floatfix]{revtex4-1}
\usepackage{graphicx}
\usepackage{epsfig}
\usepackage{amsbsy}
\usepackage{color}
\usepackage{epsfig,color}
\usepackage[utf8]{inputenc}
\usepackage{dingbat}
\usepackage{multirow}
\usepackage{amsmath}
\usepackage{array}

\newcolumntype{C}[1]{>{\centering\let\newline\\\arraybackslash\hspace{0pt}}m{#1}}

\newcommand*{\myfont}{\fontfamily{phv}\selectfont}
\newcommand*{\imaginary}{\textrm{{\myfont i}}}

\begin{document}
\title{The Coulomb-interaction-induced breaking of the Aufbau principle for hole charging of InGaAs/GaAs quantum dots}
\author{W. J. Pasek} \address{AGH University of Science and Technology,\\
Faculty of Physics and Applied Computer Science,
al. Mickiewicza 30, 30-059 Krak{\'o}w, Poland}
\date{\today}

\begin{abstract}
The so called "incomplete hole shell filling" phenomenon, that is the breaking of the Aufbau principle was reported by D. Reuter \textit{et al.}, \mbox{[Phys. Rev. Lett. \textbf{94}, 026808]} in the hole charging spectra of quantum dot when results were interpreted in context of $s$/$p$/$d$ shell system -- typical for electrons. We report an example of inter-particle-interaction induced Aufbau principle violation even if it is applied to one-particle Kohn-Luttinger eigenstates. We present a $\vec{k}{\cdot}\vec{p}$/configuration-interaction study that concerns multiple holes confined in InGaAs/GaAs self-assembled cylindrical quantum dot. Eigenenergies and eigenvectors of up to six hole ground states were obtained -- along with corresponding one-hole orbital occupations -- and discussed in context of the Aufbau principle. 
\end{abstract}

\pacs{73.21.La, 81.07.Ta, 71.70.Ej}

\maketitle

\section{Introduction}

Quantum dots are structures of size of nanometres to micrometres that can confine charge carriers (conduction band electrons and/or valence band holes) in all three directions. This kind of confinement leads to energy quantization and gives a discrete spectrum of energy levels. This is an analog of discrete spectrum of natural atoms and thus quantum dots are often called artificial atoms.\cite{wstep1,wstep2} Quantum dots of especially small size are made by applying an electrostatic potential [see Refs. \onlinecite{wstep3,wstep4,wstep5}] and by self-organization [see Refs. \onlinecite{wstep6,wstep7}]. In these very small structures confining conduction band electrons single-particle energy levels separation is so large that observation of occupation of the individual orbitals in many-electron states was possible. The sequence of the occupation for electrons in small quantum dots is generally governed by the Aufbau principle and Hund's rule outside the energy-level crossings induced by external magnetic field.\cite{wstep3,wstep4,wstep6,wstep7,wstep8}

In Ref. \onlinecite{reuter} the hole charging spectra of self-assembled InAs quantum dots in perpendicular magnetic fields were studied by capacitance-voltage spectroscopy. The authors of that work interpreted the results in the terms of the typical results obtained for electrons \textit{i.e.}~$s$, $p$ and $d$ shell system for envelope functions. From the magnetic-field dependence of the individual peaks it is concluded that the \textit{s-like} ground state is completely filled with two holes but that the fourfold degenerate \textit{p} shell is only half filled with two holes before the filling of the \textit{d} shell starts. This so called "incomplete shell filling" is attributed by them to a large influence of the Coulomb interaction in this system.

Climente \textit{et al.}\cite{planelles1} suggested that by using a model that takes into account valence band mixing via the Kohn-Luttinger (KL) Hamiltonian the behaviour of the system can be interpreted as abiding the Aufbau principle in the context of one-particle KL shells instead of the electron-type ones. However, we report an example of inter-particle-interaction induced Aufbau principle violation even if it is applied to one-particle KL eigenstates.

In this work we present a $\vec{k}{\cdot}\vec{p}$/configuration-interaction study of multiple-hole ground states of InGaAs/GaAs self-assembled cylindrical quantum dot. The model that was used allows for valence band mixing \textit{via} the 6-band Luttinger-Kohn Hamiltonian. The dot is embedded in external magnetic field applied in the growth direction (\textit{i.e.} along the symmetry axis). The lattice constant mismatch-induced strain is taken into account by the Pikus-Bir Hamiltonian.

The ground state energies and eigenstates of up to six holes confined in the system were obtained for two dot sizes: the strong and weak confinement cases. The occupation of one-particle orbitals for those wavefunctions was also calculated. We present a few cases when the Aufbau principle is evidently broken due to the Coulomb interaction and relatively small difference in energy of related one-particle levels (a "strong" violation). We also describe the so called "weak" violations of the Aufbau principle. They consist of a shift of the value of magnetic field for which a change in orbital character of a multi-hole state takes place and the value for which a corresponding level crossing occurs in the single hole spectrum. Additionally we show that both the overall value of chemical potential $\mu(N_p,B_z)$ and its relative dependence on the magnetic field cannot be inferred without considering the mentioned interaction.

\section{Theory}

\subsection{KL Hamiltonian}
We work within the envelope ansatz using $6$-band axial approximation of the Kohn-Luttinger (KL) Hamiltonian. It is written in Bloch basis of
\begin{eqnarray}
\left(\Big\vert J^{Bl}=\frac{3}{2},J^{Bl}_z=\frac{3}{2}\Big\rangle, 
\Big\vert J^{Bl}=\frac{3}{2},J^{Bl}_z=\frac{1}{2}\Big\rangle, \right. \nonumber \\
\left. \Big\vert J^{Bl}=\frac{3}{2},J^{Bl}_z=-\frac{1}{2}\Big\rangle,
\Big\vert J^{Bl}=\frac{3}{2},J^{Bl}_z=-\frac{3}{2}\Big\rangle,\right. \nonumber \\
\left. \Big\vert J^{Bl}=\frac{1}{2},J^{Bl}_z=\frac{1}{2}\Big\rangle,
\Big\vert J^{Bl}=\frac{1}{2},J^{Bl}_z=-\frac{1}{2}\Big\rangle\right),
\label{Bloch_basis} \end{eqnarray}
 where $J^{Bl}$ is the total angular momentum of the Bloch function and $J^{Bl}_z$ is its component along the symmetry ($z$) axis.\cite{annotation0} $J^{Bl}=\frac{3}{2}$ and $|J_z^{Bl}|=\frac{3}{2}$ corresponds to heavy hole bands, $J^{Bl}=\frac{3}{2}$ and $|J_z^{Bl}|=\frac{1}{2}$ corresponds to light hole components and $J^{Bl}=\frac{1}{2}$ corresponds to spin-orbit split-off bands. The Hamiltonian has the form of:
\small
\begin{widetext}
\begin{equation*}
\hat{H}_{KL}= \left(
\begin{array}{cccccc}
\hat{T}_{J^{Bl}_z=\pm\frac{3}{2}} & -\hat{S} & \hat{R} & 0 & \frac{-\hat{S}}{\sqrt{2}} & \sqrt{2}\hat{R} \\
-\hat{S}^{*} & \hat{T}_{J^{Bl}_z=\pm\frac{1}{2}} & 0 & \hat{R} & -\sqrt{2}\hat{Q} & \sqrt{\frac{3}{2}} \hat{S} \\
\hat{R}^{*} & 0 & \hat{T}_{J^{Bl}_z=\pm\frac{1}{2}} & \hat{S} & \sqrt{\frac{3}{2}} \hat{S}^{*} & \sqrt{2}\hat{Q} \\
0 & \hat{R}^{*} & \hat{S}^{*} & \hat{T}_{J^{Bl}_z=\pm\frac{3}{2}} & -\sqrt{2}\hat{R}^{*} & \frac{-\hat{S}^{*}}{\sqrt{2}} \\
\frac{-\hat{S}^{*}}{\sqrt{2}} & -\sqrt{2}\hat{Q}^{*} & \sqrt{\frac{3}{2}}\hat{S} & -\sqrt{2}\hat{R} & \hat{T}_{J^{Bl}=\frac{1}{2}}{+}{\Delta}_{SO} & 0 \\
\sqrt{2}\hat{R}^{*} & \sqrt{\frac{3}{2}}\hat{S}^{*} & \sqrt{2}\hat{Q}^{*} & \frac{-\hat{S}}{\sqrt{2}} & 0 & \hat{T}_{J^{Bl}=\frac{1}{2}}{+}{\Delta}_{SO}
\end{array}
\right),
\label{KL_rozpiska}
\end{equation*}
\end{widetext}
\normalsize
where $\hat{R} = -\frac{\sqrt{3}}{2} \frac{\gamma_{2} + \gamma_{3}}{2} \hat{p}^{2}_{-}$, $\hat{S} = \sqrt{3} \gamma_{3} \hat{p}_{-} \hat{p}_{z}$, $\hat{Q} = -\frac{\gamma_{2}}{2} [\hat{p}^{2}_{\perp} - 2 \hat{p}^2_{z}]$, $\hat{p}_{-} = \hat{p}_{x}-\imaginary\hat{p}_{y}$, $\imaginary$ is the imaginary unit and band Hamiltonians are given by:
{\small \begin{eqnarray}
\hat{T}_{J^{Bl}_z = \pm\frac{3}{2}}= -\frac{1}{2} [(\gamma_{1} + \gamma_{2}) \hat{p}^{2}_{\perp} + (\gamma_{1} - 2 \gamma_{2}) \hat{p}^2_{z}] +  V^{pot}(\vec{r}), \nonumber \\
\hat{T}_{J^{Bl}_z = \pm\frac{1}{2}}= -\frac{1}{2} [(\gamma_{1} - \gamma_{2}) \hat{p}^{2}_{\perp} + (\gamma_{1} + 2 \gamma_{2}) \hat{p}^2_{z}] +  V^{pot}(\vec{r}), \nonumber \\
\hat{T}_{J^{Bl} = \frac{1}{2}}= -\frac{\gamma_{1}}{2} [\hat{p}^{2}_{\perp} + \hat{p}^2_{z}] +  V^{pot}(\vec{r}),
\label{onebandoneparticle_KL_ham}
\end{eqnarray}}
where the in-plane envelope momentum operator is: $\hat{p}_{\perp} = \{\hat{p}_{x}, \hat{p}_{y}, 0\}$.\cite{chuang} $\gamma_1$, $\gamma_2$ and $\gamma_3$ are the Luttinger parameters -- the values for the barrier material are adopted for the whole system for simplicity.

\subsection{Basic informations on the system}
We consider a system of up to six holes confined in a cylindrical quantum dot (made of InGaAs/GaAs) in the presence of the external magnetic field. The $z$-axis is the direction of the growth and the symmetry axis. Two sizes of the dot are considered: the smaller one corresponds to strong confinement of particles, while the bigger one -- to weak confinement. In the former case the radius of the dot $R_{dot}$ is assumed to be $10$ nm and the height of the dot $2 Z_{dot}$ is $2$ nm. The values for the latter case are: $R_{dot}=20$ nm and $2 Z_{dot}=6$ nm, respectively.

Band parameters are taken from Ref. \onlinecite{bible} and correspond to $\textrm{Ga}_{1-x}\textrm{In}_{x}\textrm{As}$ for \mbox{$x=0.53$}. The Luttinger parameters for InGaAs are taken as: \mbox{$\gamma_{1} = 11.01$}, \mbox{$\gamma_{2} = 4.18$}, \mbox{$\gamma_{3} = 4.84$}. Spin-orbit splitting is assumed to be ${\Delta}_{SO} = 329.6$ meV. Valence band offset of dot material in respect to barrier material is equal to \mbox{$V_0 = -206$ meV}.  The deformation potential of dot material $a_V^{GaInAs}$ is \mbox{$678.8$ meV} and for barrier material \mbox{$a_V^{GaAs}=700$ meV}, and the deformation potential \mbox{$b_V^{GaInAs}$ is $-1894$ meV} while for barrier material \mbox{$b_V^{GaAs}=-2000$ meV}. The interpolation between values for InAs and GaAs takes into account bonding parameters, where appropriate (see Ref. \onlinecite{bible}).

The confinement potential arises from the difference between the energy of top of valence band in both materials. We set the potential energy outside of the dot to be equal to zero. In our model the confinement potential is given by a following function:
\begin{equation}
V^{pot}(\vec{r}) = V^{pot}(\rho,\phi,z) = \left\{
\begin{array}{ccc}
V_{0} &,& \rho < R_{dot} \land |z| < {Z_{dot}}\\
0 &,& \rho > R_{dot} \lor |z| > {Z_{dot}}
\end{array}
\right\}
\end{equation}

The external magnetic field is applied in the growth direction : \mbox{$\vec{B} = (0,0,B_z)$}.

\subsection{Strain effects}
The difference between lattice constants of dot material and of barrier material \mbox{($\epsilon_0 = \frac{a_{GaInAs}}{a_{GaAs}} - 1 = 3.8\%$)} is the source of stress and strain in the system. We use elastic model of Ref. \onlinecite{davies} to calculate the strain field. The relative strain is given by:
\begin{equation}
\epsilon_{ij}^{rel}(\vec{r}) = -\frac{\epsilon_0}{4 \pi} \frac{1+\nu}{1-\nu} \oint \frac{(\vec{r}_i-{\vec{r^{'}_i}}) dS_j^{'}}{{|r-r^{'}|}^3},
\end{equation}
where the Poisson's ratio $\nu = \frac{1}{3}$, $(i,j) \in \{x,y,z\}\otimes\{x,y,z\}$ and the integration is conducted over the surface of the dot. The hydrostatic strain is given by $\epsilon_{ii} = \epsilon_{0}$ inside the dot and is equal to zero outside.

The effect of this strain on the energy of the system is given by the Pikus-Bir Hamiltonian.\cite{chuang} However, we restrict ourselves to biaxial strain only (as in Ref. \onlinecite{planelles2}) by assuming: $\epsilon_{xx}=\epsilon_{yy} \neq \epsilon_{zz}$ and $\epsilon_{xy}=\epsilon_{yz}=\epsilon_{zx} = 0$. Written in the Bloch basis of \mbox{Eq. \ref{Bloch_basis}} the biaxial Pikus-Bir Hamiltonian has a form of:
\begin{widetext}
\begin{equation}
\hat{H}_{PB}= \left(
\begin{array}{cccccc}
\hat{P}_{\epsilon}+\hat{Q}_{\epsilon} & 0 & 0 & 0 & 0 & 0 \\
0 & \hat{P}_{\epsilon}-\hat{Q}_{\epsilon} & 0 & 0 & -\sqrt{2}\hat{Q}_{\epsilon} & 0 \\
0 & 0 & \hat{P}_{\epsilon}-\hat{Q}_{\epsilon} & 0 & 0 & \sqrt{2}\hat{Q}_{\epsilon} \\
0 & 0 & 0 & \hat{P}_{\epsilon}+\hat{Q}_{\epsilon} & 0 & 0 \\
0 & -\sqrt{2}\hat{Q}_{\epsilon} & 0 & 0 & \hat{P}_{\epsilon} & 0 \\
0 & 0 & \sqrt{2}\hat{Q}_{\epsilon} & 0 & 0 & \hat{P}_{\epsilon}
\end{array}
\right),
\label{Pikus-Bir}\end{equation}
\end{widetext}
where $\hat{P}_{\epsilon} = -a_v (\epsilon_{xx}+\epsilon_{yy}+\epsilon_{zz})$ and $\hat{Q}_{\epsilon}= \frac{b_v}{2} (\epsilon_{xx} + \epsilon_{yy} - 2 \epsilon_{zz})$.\cite{annotation1}

\subsection{Magnetic field}\label{subsec:Magnetic field}
There are several propositions how to include external magnetic field $B_z$ into the KL model. The impact of the field on the envelope functions can be taken into account by substituting the canonical momentum $\hat{p}=-\imaginary\nabla$ by $\hat{p}=-\imaginary\nabla-\vec{A}$ in Eq.(\ref{KL_rozpiska}) like \textit{e.g.} in Ref. \onlinecite{regohawrylak}. Planelles and Jaskólski suggested reversing the order of operations: to start with considering magnetic field in context of each band separately and then subsequently to apply the envelope approximation in $\vec{k}{\cdot}\vec{p}$ procedure.\cite{planelles0} This approach leads to different magnetic terms in KL Hamiltonian, specifically that all quadratic terms are diagonal and - in 4-band model - also the linear terms are. Further work by Climente \mbox{\textit{et al.}} compared both approaches in scope of 4-band KL study of hole in InAs/GaAs quantum molecules.\cite{planelles2} Authors concluded that the second method gives results that are in agreement with experiment while the first one artificially enhances the HH-LH mixing (by means of off-diagonal terms) which leads to not observed bonding-antibonding ground state magnetic switching. Work Ref. \onlinecite{planelles2} also refined the second model by putting the relevant effective masses for direction perpendicular to the growth axis into magnetic terms instead of the effective masses in direction of that axis. After this change the model correctly retrieves single-band limit in case of band-decoupling. Finally, the model was reshaped once more in Ref. \onlinecite{planelles3} by including spin degree of freedom and by defining Zeeman terms for Bloch functions \textit{via} hole $g$-factors, independently of relevant Zeeman envelope-dependent ones. The latter work predicts a quadratical increase of the excitonic gap with increasing magnetic field, as observed in photoluminescence experiments of InGaAs QDs.

In our work we follow Ref. \onlinecite{planelles3} in including the magnetic field $B_z$ into the model as:
{\small \begin{eqnarray}
(\hat{H}_{B_z})_{11} = (\gamma_1+\gamma_2) \left(\frac{B_z^2 \rho^2}{8}+\frac{B_z}{2} J_z^{env}\right) + \frac{3}{2} \kappa \mu_B B_z \nonumber \\
(\hat{H}_{B_z})_{22} = (\gamma_1-\gamma_2) \left(\frac{B_z^2 \rho^2}{8}+\frac{B_z}{2} J_z^{env}\right) + \frac{1}{2} \kappa \mu_B B_z \nonumber \\
(\hat{H}_{B_z})_{33} = (\gamma_1-\gamma_2) \left(\frac{B_z^2 \rho^2}{8}+\frac{B_z}{2} J_z^{env}\right) - \frac{1}{2} \kappa \mu_B B_z \nonumber \\
(\hat{H}_{B_z})_{44} = (\gamma_1+\gamma_2) \left(\frac{B_z^2 \rho^2}{8}+\frac{B_z}{2} J_z^{env}\right) - \frac{3}{2} \kappa \mu_B B_z \nonumber \\
(\hat{H}_{B_z})_{55} = \gamma_1 \left(\frac{B_z^2 \rho^2}{8}+\frac{B_z}{2} J_z^{env}\right) + \frac{1}{2} \kappa^{'} \mu_B B_z \nonumber \\
(\hat{H}_{B_z})_{66} = \gamma_1 \left(\frac{B_z^2 \rho^2}{8}+\frac{B_z}{2} J_z^{env}\right) - \frac{1}{2} \kappa^{'} \mu_B B_z \nonumber \\
(\hat{H}_{B_z})_{i{\ne}j} = 0,
\label{HBZ} \end{eqnarray}}
where $J_z^{env}$ is the total angular momentum of envelope, $\mu_B$ is Bohr magneton and $\kappa=\frac{4}{3}$, $\kappa^{'}=\frac{2}{3}$ are effective hole $g$-factors.

\subsection{Computation process}
The first step in the computation process is to solve eigenproblem of the $\hat{H}_{KL}+\hat{H}_{PB}$ for a single hole. The variational basis for envelope contains functions of the following kind: 
{\scriptsize \begin{eqnarray}
\psi_{k,J_z^{env},n}(\vec{r}) = \frac{\exp(\imaginary J_z^{env} \phi)}{\sqrt{2 \pi}} BF_{k,J_z^{env}}\left(\frac{\rho K_{k,{J_z^{env}}}}{R_{ef}}\right) ZF_{n}\left(\frac{z}{Z_{ef}}\right)
\label{onebandoneparticle_wf} \end{eqnarray}}
where $BF_{k,J_z^{env}}$ is a relevant normalized Bessel J function ($K_{k,J_z^{env}}$ is its $k$-th zero) for $\rho < R_{ef}$ and zero otherwise. The function $ZF_{n}$ has the form of:
{\small \begin{equation}
ZF_{n}(y)=\frac{1}{\sqrt{Z_{ef}}} \cos\left(\frac{\pi}{2}[n (y - 1) + 1 + (n-1) (n~\textrm{mod}~2)]\right),
\end{equation}} for $|z| < {Z_{ef}}$ and zero otherwise. The $R_{ef}$ is an effective radius of the wavefunction, that is assumed to be not less than the radius of the dot confinement potential: $R_{ef} \geq R_{dot}$. It is given by $R_{ef} = \frac{R_{dot}}{b_{R}}$. Similarly, the $Z_{ef} \geq {Z_{dot}}$ is the effective half-height of the dot, and $Z_{ef} = \frac{Z_{dot}}{a_{HH}}$ in case of heavy-hole bands and $Z_{ef} = \frac{Z_{dot}}{a_{LHSO}}$ for light-hole and split-off bands. 

The $z$-component of total angular momentum is defined for a KL eigenfunction and is equal to the sum of the total angular momentum of the Bloch state and the angular momentum of envelope $z$-components: $J_z = J_z^{env} + J_z^{Bl}$ (\textit{i.e.} the axial approximation of KL Hamiltonian commutes with the operator of the total angular momentum of the hole). This allows for very significant simplification of the variational basis. The computation can be done for each $J_z$ separately and only one value of $J_z^{env} = J_z - J_z^{Bl}$ is used for each component of KL state vector in each of these calculations. This fact was taken advantage of in case of dots with axial symmetry in numerous works (see \textit{e.g.} \onlinecite{planelles1, regohawrylak, abhgst, abhgst2, abhgst3, abhgst4, 2h-1dot-calc-3} among many others).

We considered $J_z$ values from the range $\{-\frac{9}{2},...,\frac{9}{2}\}$. The $k$ and $n$ quantum numbers numerate the basis functions in $\rho$ and $z$ directions, respectively. The set of values that was used in calculation is $k\in\{1,8\}$ and $z\in\{1,6\}$. 

At this stage $\{a_{HH}, a_{LHSO}, b_{R}\}$ constitutes the set of the variational parameters. Each element of this set can in principle have any value in range of $(0,1]$. We conducted calculations for each case when parameter is equal to one the values: $\frac{j}{20}$ for $j{\in}\{1,19\}$ ($j$-s of parameters are independent). For the case of the strong confinement the lowest KL ground-state energy was obtained in case of the following set of values: $\{a_{HH}=0.25,a_{LHSO}=0.40,b_{R}=0.55\}$, what is equivalent to: $R_{ef}=1.82 R_{dot}$, $Z_{ef} = 4 Z_{dot}$ in the case of heavy hole bands and $Z_{ef} = 2.5 Z_{dot}$ for light hole and split-off bands. By an analogical procedure for the bigger dot following values were obtained: $\{a_{HH}=0.40,a_{LHSO}=0.55,b_{R}=0.75\}$ and consequently $R_{ef}=1.33 R_{dot}$, $Z_{ef} = 2.5 Z_{dot} / 1.82 Z_{dot}$ for heavy holes / light and split-off holes, respectively.

The matrix elements of off-diagonal operators $\hat{R},\hat{S}$ of Eq. \ref{KL_rozpiska} are calculated analytically where possible and otherwise by Legendre-Gauss quadratures with $1000$ points. All other matrix elements of this Hamiltonian can be obtained analytically. The calculation of matrix elements of Pikus-Bir Hamiltonian (Eq. \ref{Pikus-Bir}) was conducted by rectangle integration on a mesh with $dz = 0.1$ nm and $d\rho = 0.2$ nm in the case of the smaller dot, and two times bigger spacings for the weak confinement system. In order to obtain relevant strain tensor elements we used the same method on a mesh with $d\phi^{'} = \frac{\pi}{1024}$.

Resulting eigenvectors of envelope have the form of:
\begin{equation}
\phi_{J_z,m_{J_z}}(\vec{r}) =
\left(
\begin{array}{c}
\xi^{J^{Bl}_z=\frac{3}{2}}_{m_{J_z}}(\rho,z) e^{{\imaginary} (J_z-\frac{3}{2}) \phi} \\
\xi^{J^{Bl}_z=-\frac{1}{2}}_{m_{J_z}}(\rho,z) e^{{\imaginary} (J_z+\frac{1}{2}) \phi} \\
\xi^{J^{Bl}_z=\frac{1}{2}}_{m_{J_z}}(\rho,z) e^{{\imaginary} (J_z-\frac{1}{2}) \phi} \\
\xi^{J^{Bl}_z=-\frac{3}{2}}_{m_{J_z}}(\rho,z) e^{{\imaginary} (J_z+\frac{3}{2}) \phi}
\end{array}
\right),
\label{ho1peigen}
\end{equation}
with definite total angular momentum of $J_z$, where $\xi^{J^{Bl}_z}_{m_{J_z}}$ is one-band wavefunction component and $m_{J_z}~\in~\{1,2,3,...\}$ numerates states in order of increasing energy for a given $J_z$. The states $\phi_{J_z,m}$ and $\phi_{-J_z,m_{-J_z}}$ are degenerate for all $J_z$ and $m_{J_z}$ when no external magnetic field $B_z$ is applied to the system. These eigenfunctions are used for construction of basis states for many-hole calculation.

In the special case of one hole it is easy to include the magnetic field in the computation directly. The eigenstates of $\hat{H}_{KL}+\hat{H}_{PB}+\hat{H}_{B_z}$ are obtained in the same basis as described before. The matrix elements of Eq. \ref{HBZ} are calculated analytically when possible and by Legendre-Gauss quadratures with 1000 points otherwise (some cases of "diamagnetic" components). Note that adding the magnetic field in the $z$ direction does not affect the axial symmetry of the system and $J_z$ is still defined. As a consequence the eigenstates of $\hat{H}_{KL}+\hat{H}_{PB}+\hat{H}_{B_z}$ can also be written in the form of Eq. \ref{ho1peigen}.

In the case of many hole states the variational basis is constructed from the KL eigenfunctions obtained for the case of no magnetic field. After that the full many-body Hamiltonian (that is the single-hole energies plus magnetic and Coulomb terms) is diagonalized in this basis. The total angular momentum $z$-component of many-hole state is defined:
\begin{equation}
J(N_P)H_z = \sum_{i_P=1}^{N_P} [J_z]_{i_P},
\end{equation}
where index $i_P$ runs over all particles and $N_P$ is the number of particles in a given case. Thus the construction of the basis for each $J(N_P)H_z$ (and the diagonalization following it) can be done separately. There are two parameters that describe this process: $NV_A$ and $NV_B$. At the beginning we find all the sets of particular $J_z$-s that sum to $J(N_P)H_z$. For each sequence $\left([J_z]_1,...,[J_z]_{N_P}\right)$ and for each particle we take into account $m_{J_z}=1,..,NV_A$ one-particle eigenfunctions and form sequences $\left(\phi_{[J_z]_{1},m_{[J_z]_{1}}},...,\phi_{[J_z]_{N_P},m_{[J_z]_{N_P}}}\right)$. Multiple equivalent sequences are reduced to a single instance.\cite{annotation2} Secondly, we sort all the latter-type sequences by the sum of respective one-particle energies. Then we take $NV_B$ sequences of lowest sums and construct the relevant Slater determinants:
\begin{equation}
\Psi_{J(N_P)H_z,M_{J(N_P)H_z}}=\hat{A}\left(\phi_{[J_z]_{1},m_{[J_z]_{1}}},...,\phi_{[J_z]_{N_P},m_{[J_z]_{N_P}}}\right),
\end{equation}
where $\hat{A}$ is the normalized antisymmetrization operator and $M_{J(N_P)H_z}$ numerates the functions by the sum of one-particle energies. The set of all such Slater determinants is the variational basis for a given $J(N_P)H_z$.

\subsection{Coulomb interaction}
The Hamiltonian that describes Coulomb interaction between the particles $i$ and $j$ is $\hat{H}_{int}^{i,j} = \frac{1}{\varepsilon |\vec{r_i}-\vec{r_j}|}$ and the total interaction Hamiltonian is given by: $\hat{H}_{int} = \sum_{i{\neq}j} \hat{H}_{int}^{i,j}$. We adopted the electric permittivity constant in whole system equal to the value for dot material: $\varepsilon = \varepsilon_R \varepsilon_{0} = 13.9 \varepsilon_{0}$. The Coulomb matrix element of two Slater determinants can be expressed as the sum of matrix elements between the products of pair of respective wavefunctions multiplied by relevant coefficients:
{\small \begin{widetext}
\begin{eqnarray}
\langle \Psi_{J(N_P)H_z,A} \mid \hat{H}_{int} \mid \Psi_{J(N_P)H_z,B} \rangle &=& \sum_{i{\neq}j}^{N_P} C_{i,j}~I_{i,j}^{A,B} \nonumber \\ &=& \sum_{i{\neq}j}^{N_P} C_{i,j} \langle \phi_{[J_z]_{i}^{A},m_{[J_z]_{i}^{A}}} \phi_{[J_z]_{j}^{A},m_{[J_z]_{j}^{A}}} \mid \hat{H}_{int}^{i,j} \mid \phi_{[J_z]_{i}^{B},m_{[J_z]_{i}^{B}}} \phi_{[J_z]_{j}^{B},m_{[J_z]_{j}^{B}}} \rangle,
\end{eqnarray}
\end{widetext}} where the value of coefficient $C_{i,j}$ can be one of $\{-1,0,1\}$, depending on the parity of the relevant antisymetrization permutations and the values of relevant Kronecker deltas. The $A$ and $B$ indices belong to the set that contains all considered $M_{J(N_P)H_z}$ for a given $J(N_P)H_z$. The matrix elements of the right hand side may have nonzero values only if $[J_z]_{i}^{A}+[J_z]_{j}^{A}=[J_z]_{i}^{B}+[J_z]_{j}^{B}$ and could be calculated directly, by six-dimensional integration. However, it is much advantageous to translate the issue in question to three dimensional integration over an effective potential:
{\small \begin{eqnarray}
I_{i,j}^{A,B} & = & \int  \phi^{*}_{{[J_z]_j^A},m_{[J_z]_j^A}}(\vec{r}_j)~\phi_{{[J_z]_j^B},m_{[J_z]_j^B}}(\vec{r}_j)~V_{i}^{A,B}(\vec{r}_{j}) d\vec{r}_{j}  \nonumber \\
V_{i}^{A,B}(\vec{r}_{j}) &=& \int \phi^{*}_{{[J_z]_i^A},m_{[J_z]_i^A}}(\vec{r_{i}})~\hat{H}_{int}^{i,j}~\phi_{{[J_z]_i^B},m_{[J_z]_i^B}}(\vec{r_{i}}) d\vec{r}_{i}.
\end{eqnarray}}
This potential is obtained by solving the Poisson equation with complex right hand side:
\begin{equation}
\nabla^2 V_{i}^{A,B}(\vec{r}) = - \frac{4\pi}{\varepsilon_{R}} \phi^{*}_{{[J_z]_i^A},m_{[J_z]_i^A}}(\vec{r}) \phi_{{[J_z]_i^B},m_{[J_z]_i^B}}(\vec{r}).
\label{poisson}\end{equation}
In the latter task we employ multigrid approach and overrelaxation with the relaxation parameter $\omega=1.9$ and \mbox{$dz = d\rho = 0.1$ nm} on the final mesh. For a more detailed description of this computation scheme see \textit{e.g.} our earlier work Ref. \onlinecite{triony_uj}. Number of iterations used in the process is adaptive, assuring after the last iteration a sufficiently small value of inaccuracy parameter:
\begin{equation}
\int \left| \nabla^2 V_{i}^{A,B}(\vec{r}) + \frac{4\pi}{\varepsilon_{R}} \phi^{*}_{{[J_z]_i^A},m_{[J_z]_i^A}}(\vec{r}) \phi_{{[J_z]_i^B},m_{[J_z]_i^B}}(\vec{r}) \right| d\vec{r}.
\end{equation}

After the diagonalization of the total Hamiltonian of the system (\textit{i.e.} kinetic + magnetic + Coulomb) multi-hole levels are described in the form of ($J(N_P)H_z$, $m_{J(N_P)H_z}$), where $m_{J(N_P)H_z}$ numerates levels of a given $J(N_P)H_z$ in order of increasing energy.

\section{Results}
\subsection{Strong confinement}
In this section the results for the weak confinement case \textit{i.e.} the system with $R_{dot}=10$ nm and $2 Z_{dot}=2$ nm are presented.

\subsubsection{Single hole}
\begin{figure*}[!ht]
\makebox[\textwidth][c]{\begin{tabular}{lcl}
(a) &~~~& (b)\\
\rotatebox{0}{\epsfxsize=70mm \epsfbox[65 35 705 505] {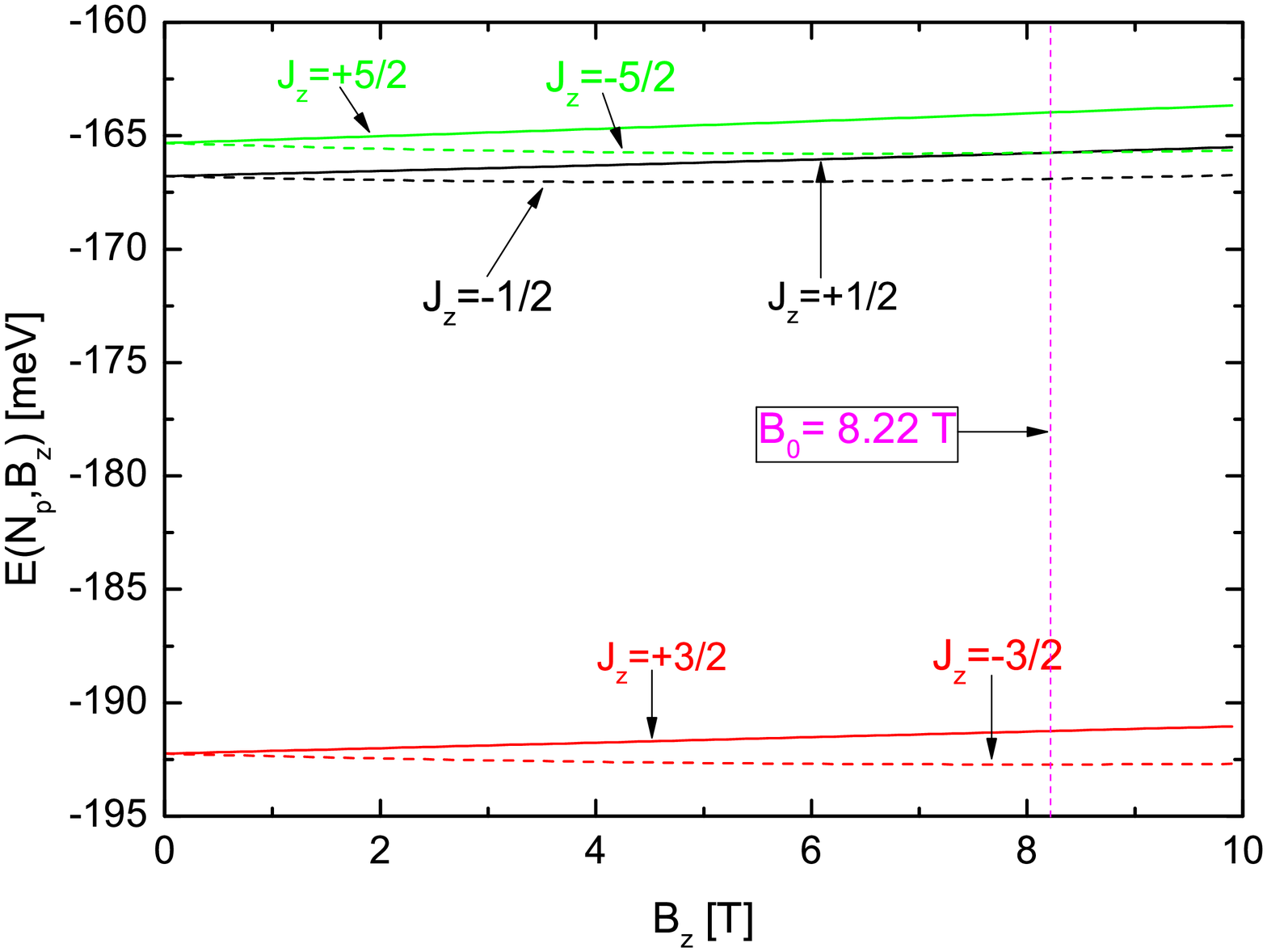}} &~~~&
\rotatebox{0}{\epsfxsize=70mm \epsfbox[65 35 705 505] {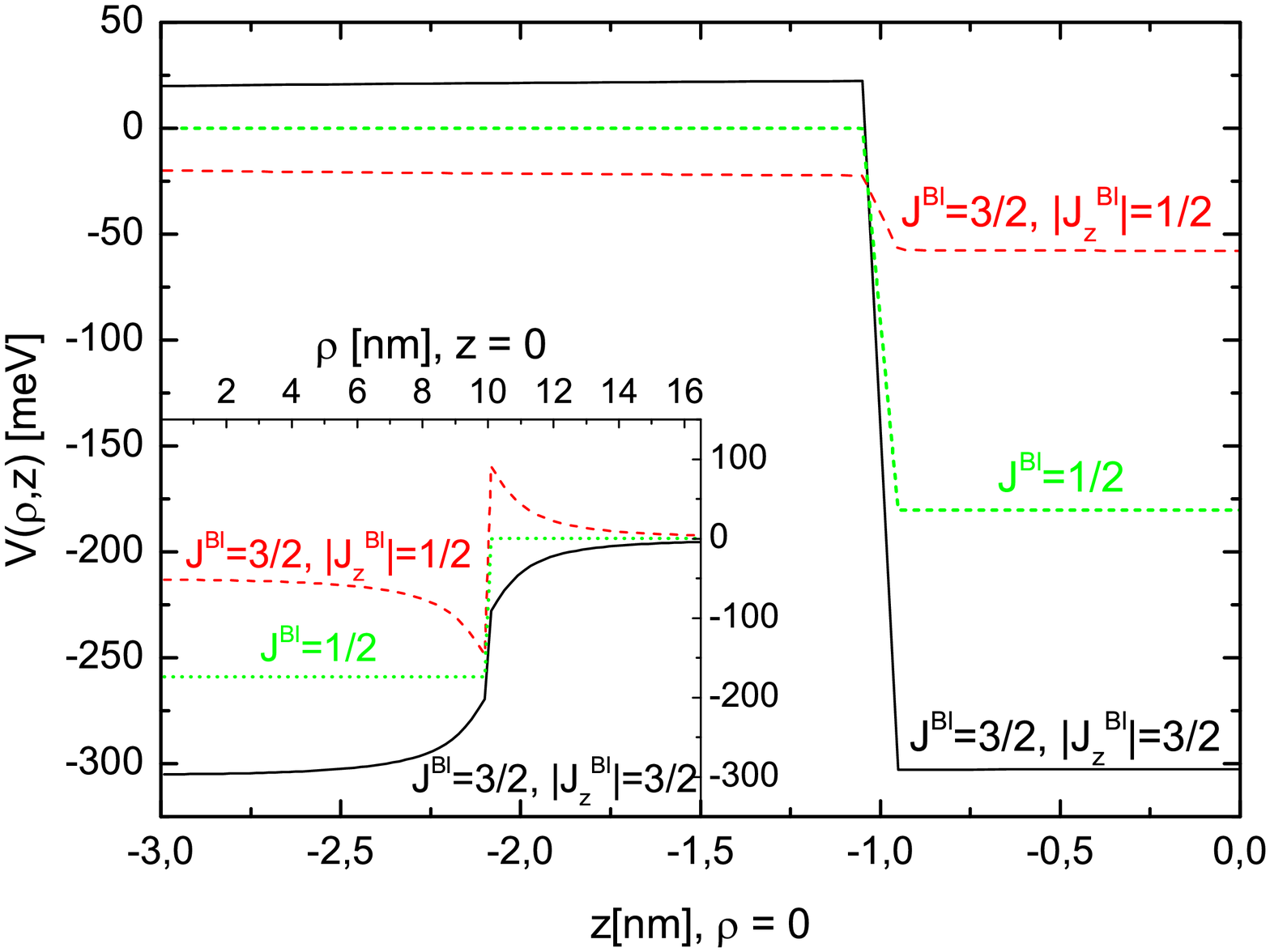}} \\
\end{tabular}}
\caption{(a) The energy spectrum of one hole in external magnetic field $B_z$; strong confinement. (b) The profile of confining potential along $z$-axis including biaxial strain for different valence bands. Inset: the profile for $z=0$ along the $x$-$y$ plane radius~$\rho$.\label{onehole}}\end{figure*}

The energy spectrum of a single hole in external magnetic field $B_z$ is presented if Fig.~\ref{onehole}(a). The profile of effective confinement potential including strain along the growth axis is shown in Fig.~\ref{onehole}(b). The inset shows the potential profile for $z=0$ \textit{i.e.} in the centre of the dot in growth direction as a function of radius in $x$-$y$ plane. As noted before, strain essentially reinforces the confinement for heavy holes ($322$ meV instead of $206$ meV without strain) and essentially weakens it for light holes ($38$ meV in the middle of the dot instead of $206$ meV without strain) while the potential for split-off bands remains nearly unaffected ($180$ meV). As one can see in Fig.~\ref{onehole}(a) -- for low magnetic field -- the six energy levels of lowest energy are (in order) states with: ($J_z=-\frac{3}{2}, m_{-\frac{3}{2}}=1$), ($J_z=\frac{3}{2}, m_{\frac{3}{2}}=1$), ($J_z=-\frac{1}{2}, m_{-\frac{1}{2}}=1$), ($J_z=\frac{1}{2}, m_{\frac{1}{2}}=1$), ($J_z=-\frac{5}{2}, m_{-\frac{5}{2}}=1$), ($J_z=\frac{5}{2}, m_{\frac{5}{2}}=1$). The above sequence changes in scope of the crossing that takes place at $B_0$ and for stronger magnetic fields the 
 ($J_z=-\frac{5}{2}, m_{-\frac{5}{2}}=1$) is the fourth level and ($J_z=\frac{1}{2}, m_{\frac{1}{2}}=1$) the fifth one in order of increasing energy.

When no magnetic field is present in the system two levels with the same $|J_z|$ are Kramer degenerate. In this case the energy separation between the ground state energy level ($|J_z|=\frac{3}{2}$) and states with $|J_z|=\frac{1}{2}$ is relatively large and equal to \mbox{$25.5$ meV}. On the other hand the energy separation between the energy level corresponding to $|J_z|=\frac{1}{2}$ and the states with $|J_z|=\frac{5}{2}$ is \mbox{$1.5$ meV} which is a small value. Also, the mentioned six levels are quite far from all other states in energy scale -- the next level (which is $J_z=-\frac{1}{2}$, $m_{-\frac{1}{2}}=2$) is $22$ meV higher. These facts bear essential consequences to the multi-particle spectra that will be discussed in later part of our work.

If a small magnetic field is introduced into the system then the response of the energy levels is mainly linear, as the diamagnetic term is significantly smaller than orbital/spin Zeeman ones. For all six levels presented in Fig.~\ref{onehole}(a) the heavy hole band of least envelope angular momentum is strongly dominating. For states with positive $J_z$ it is band with $J_z^{Bl}=\frac{3}{2}$ and for states with negative $J_z$ it is band with $J_z^{Bl}=-\frac{3}{2}$. The corresponding spin Zeeman terms are: $\frac{3}{2} \kappa \mu_B B_z$ and $-\frac{3}{2} \kappa \mu_B B_z$ which is the reason why negative $J_z$ is lower in energy in each $|J_z|$ pair. Moreover, the absolute value of spin Zeeman terms of six states are all nearly the same. However, the envelope angular momentum of the dominating band $J_z^{env}=J_z-J_z^{Bl}$ is different for each state, as presented in Table \ref{TabF1H}. In the case of states with $|J_z|=\frac{3}{2}$ dominating $J_z^{env}$ is zero and the corresponding orbital Zeeman term is also zero. For the states with $|J_z|=\frac{1}{2}$ dominating $J_z^{env}$ has sign opposite to $J_z^{Bl}$ so the orbital Zeeman term partially suppress the larger spin one. Conversely, states with $|J_z|=\frac{5}{2}$ have their respective dominating $J_z^{env}$ sign parallel to $J_z^{Bl}$ and two Zeeman terms add together. As a result the splitting between the energy levels in $|J_z|$ pair is smallest in the case of $|J_z|=\frac{1}{2}$, medium for $|J_z|=\frac{3}{2}$ and largest in the $|J_z|=\frac{5}{2}$ case -- see the values in Table~\ref{TabF1H}.

\begin{table*}[ht]
\begin{center}
    $\begin{array}{c|c|c|c|c} \hline
        \textrm{state}  & ~~J_z~~ & \textrm{dominating}~J_z^{Bl} & \textrm{dominating}~J_z^{env} & \textrm{energy splitting [meV/T]}\\ \hline
        \textrm{ground} & -3/2 & -3/2 & 0 & \multirow{2}{*}{0.234}\\
        1\textrm{st excited} & 3/2 & 3/2 & 0 \\ \hline
        2\textrm{nd excited} & -1/2 & -3/2 & 1 & \multirow{2}{*}{0.211} \\
        3\textrm{rd excited} & 1/2 & 3/2 & -1 \\ \hline
        4\textrm{th excited} & -5/2 & -3/2 & -1 & \multirow{2}{*}{0.286} \\
        5\textrm{th excited} & 5/2 & 3/2 & 1 \\ \hline   
    \end{array}$
\end{center}
\caption{Summary of the six lowest in energy single-hole states.}\label{TabF1H}\end{table*}

In this kind of systems the occupation of single-particle orbitals in multi-electron wavefunctions obey Aufbau principle. This principle states that particles occupy one-particle orbitals in order of increasing energy. If our hole system was governed by this rule, the ground state of two-hole case would be well described by assuming that the ground state and the first excited state of Fig.~\ref{onehole}(a) are occupied. In the case of three holes, it would mean that additionally the orbital of second excited state is occupied. In case of such a system we can infer the total angular momentum of a multi-hole ground state from the one-particle spectrum as:
\begin{equation}
J(N_P)H_z^{GS} = \sum_{i}^{N_P} (J_z)_{i},
\end{equation}
where $(J_z)_{i}$ is total angular momentum of one hole state that is $i$-th lowest in energy. For our system we would have: $J2H_z=0$, $J3H_z=-\frac{1}{2}$, $J4H_z=0$, $J5H_z=-\frac{5}{2}$ and $J6H_z=0$. The Aufbau principle is trivially fulfilled when no particle-particle interaction is present.

\subsubsection{Energy spectra of multiple holes}\label{subsubsec:SC-Energy-spectra-of-multiple-holes}
\begin{figure*}[!ht]
 \makebox[\textwidth][c]{
\begin{tabular}{lcl}
(a) &~~~& (b)\\
\rotatebox{0}{\epsfxsize=70mm \epsfbox[65 35 705 505] {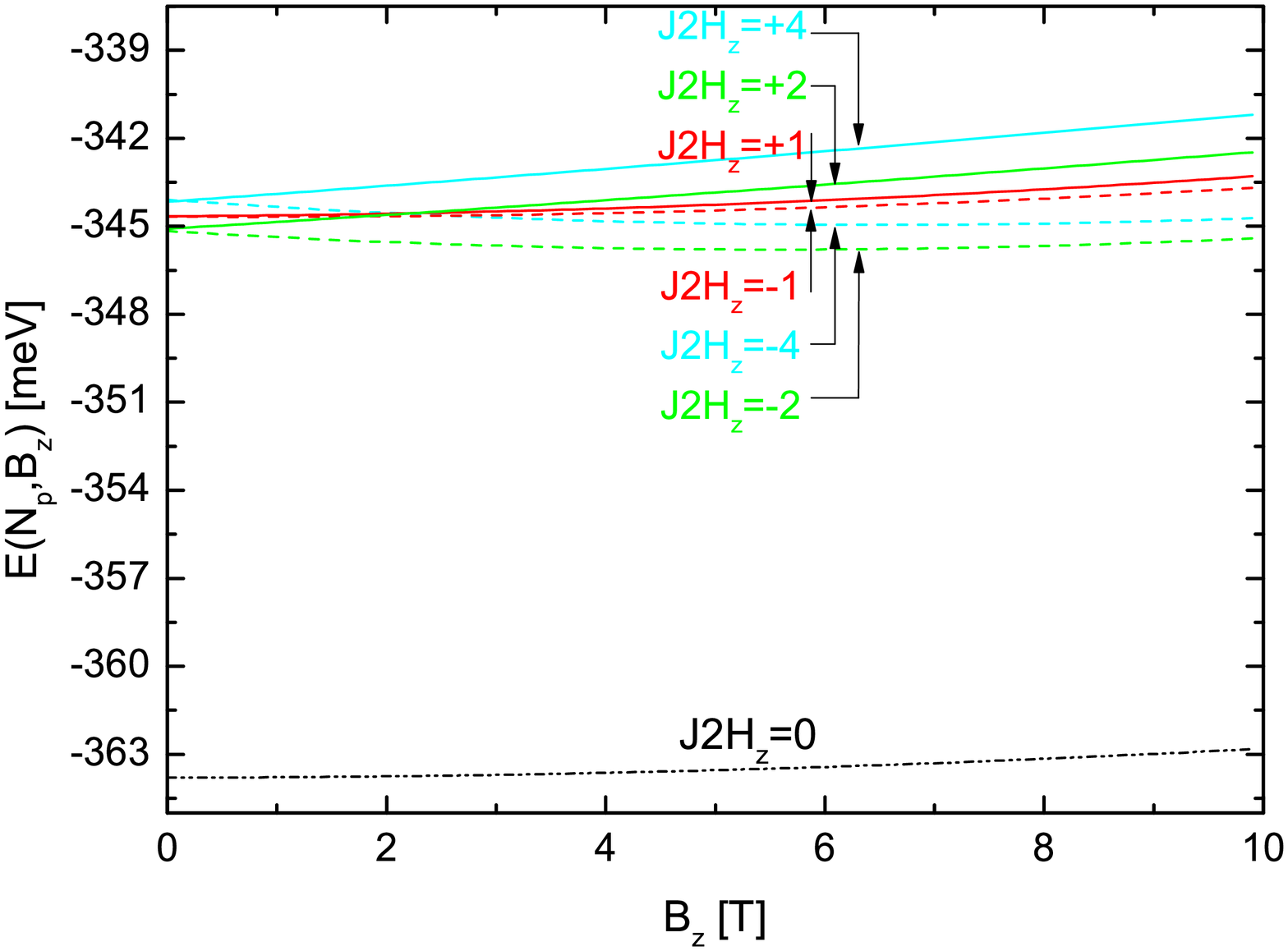}} &~~~&
\rotatebox{0}{\epsfxsize=70mm \epsfbox[65 35 705 505] {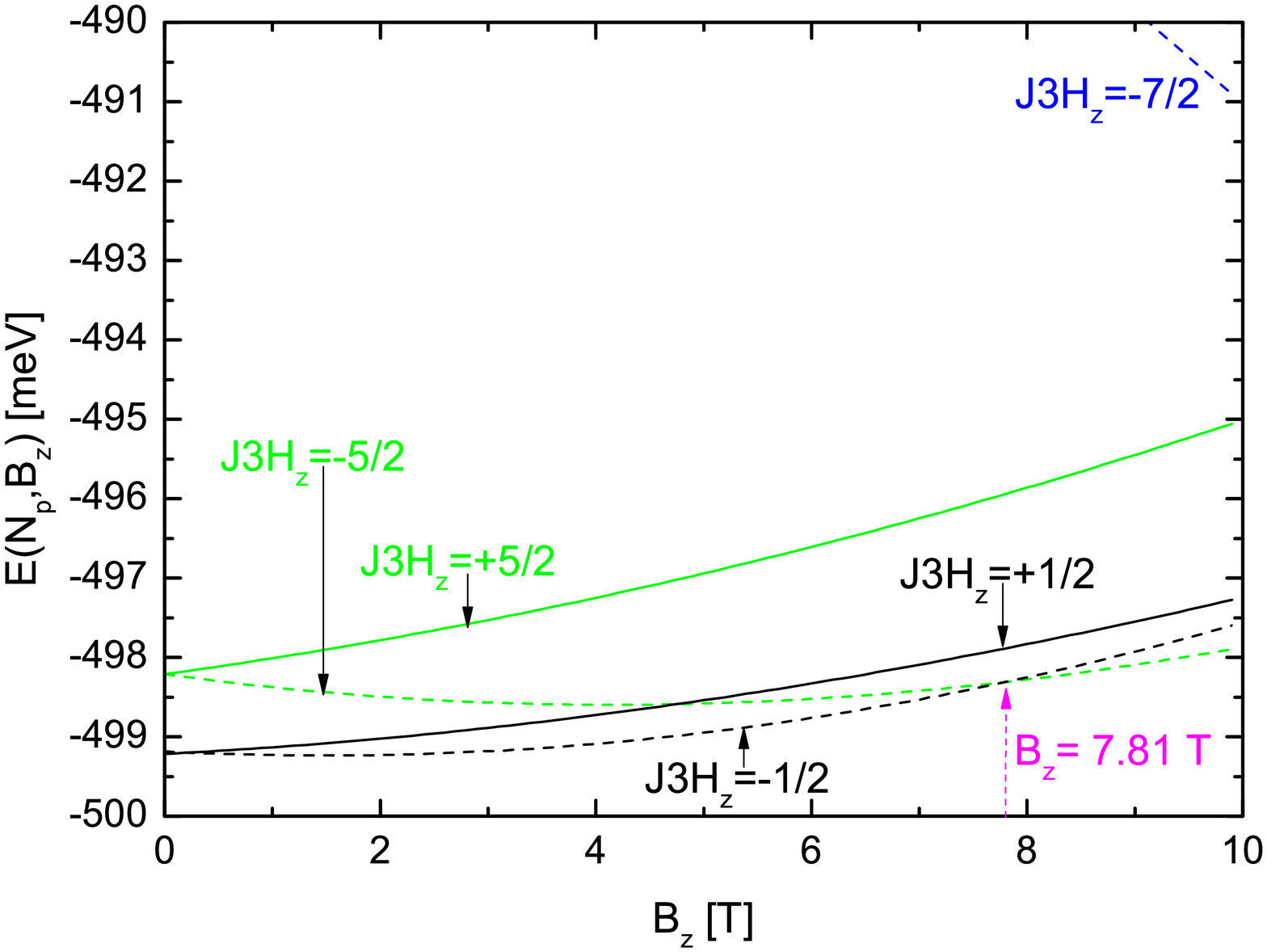}} \\
(c) &~~~& (d)\\
\rotatebox{0}{\epsfxsize=70mm \epsfbox[65 35 705 505] {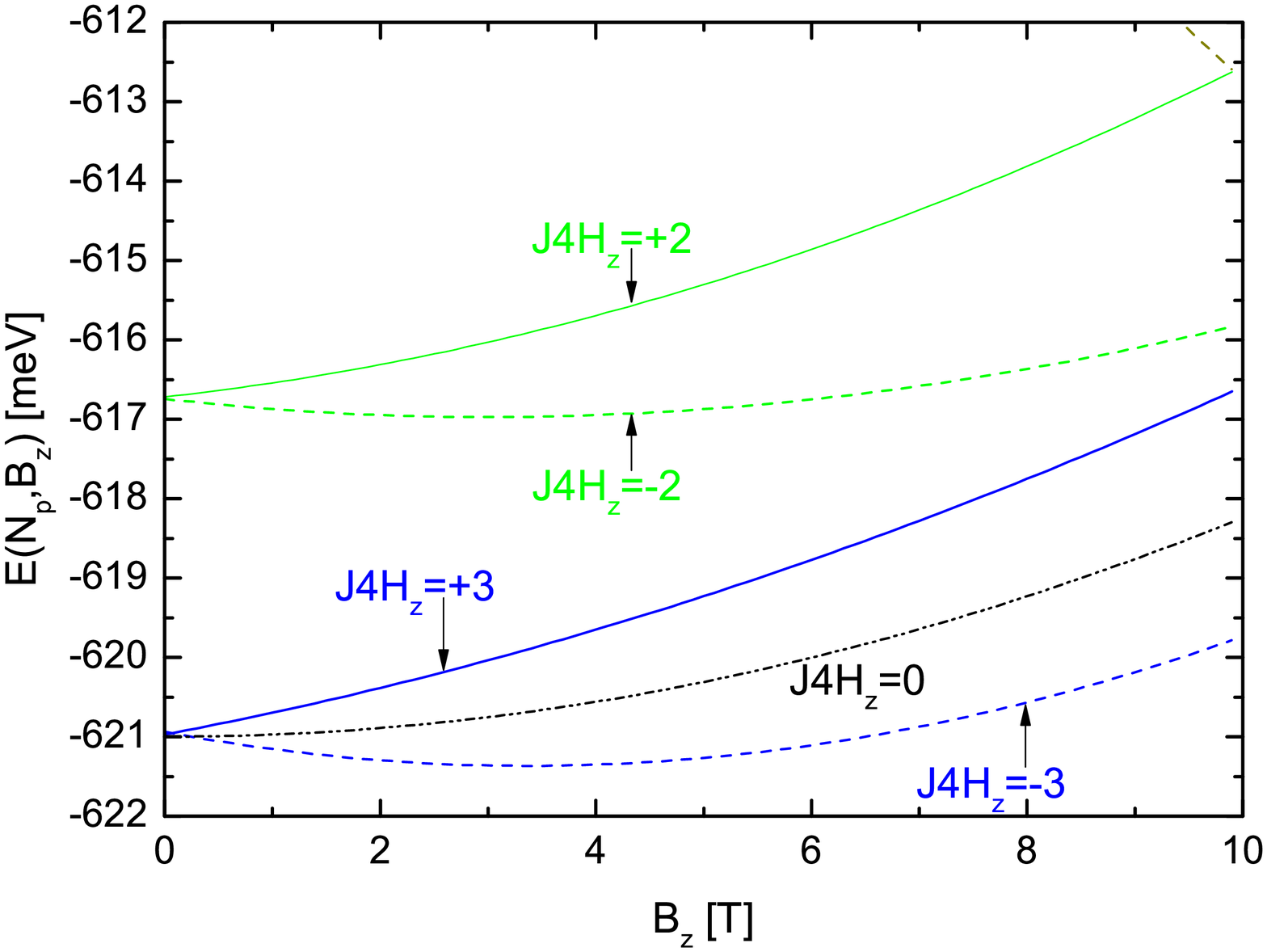}} &~~~&
\rotatebox{0}{\epsfxsize=70mm \epsfbox[65 35 705 505] {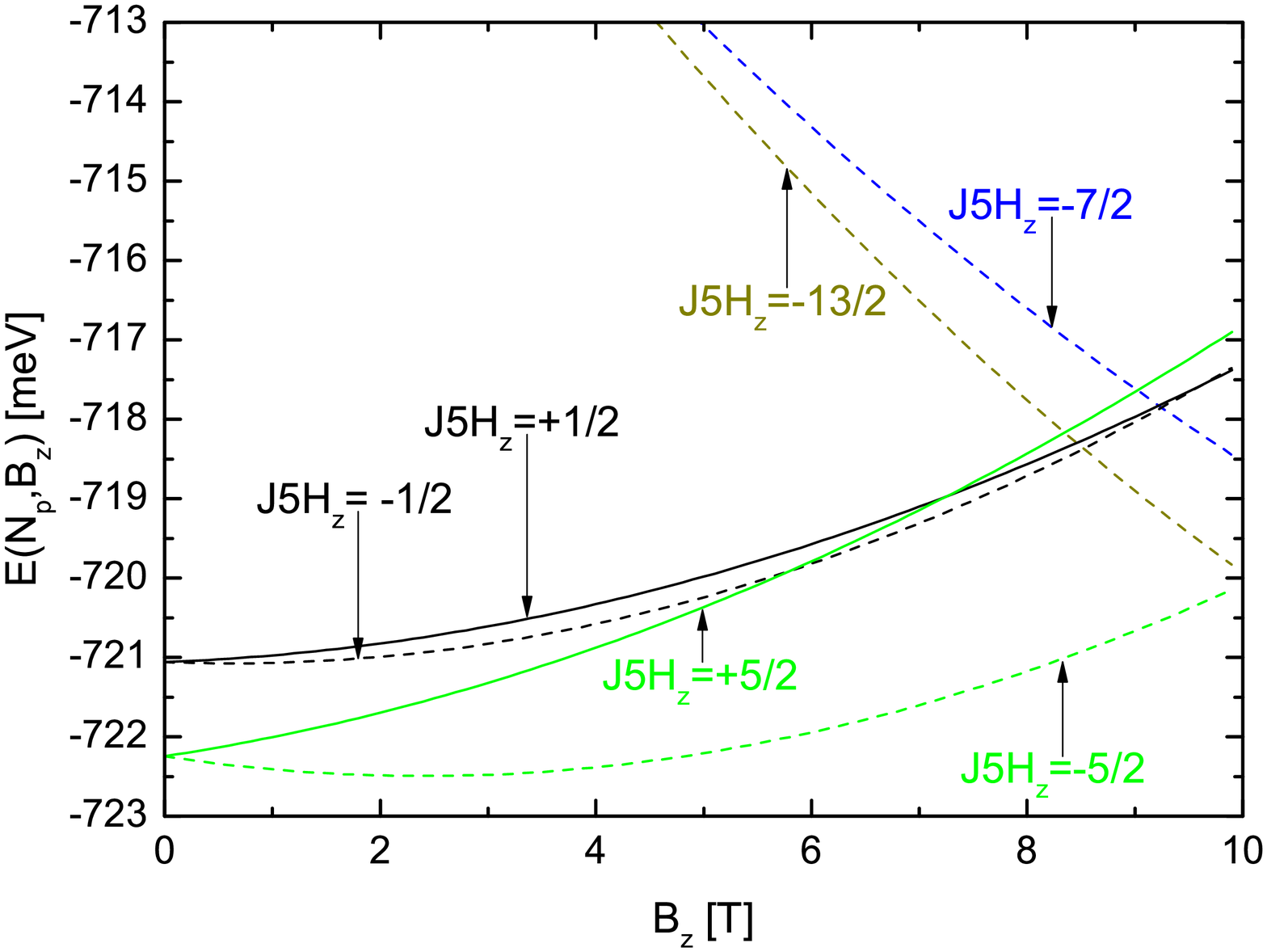}} \\
(e) &~~~& (f)\\
\rotatebox{0}{\epsfxsize=70mm \epsfbox[65 35 705 505] {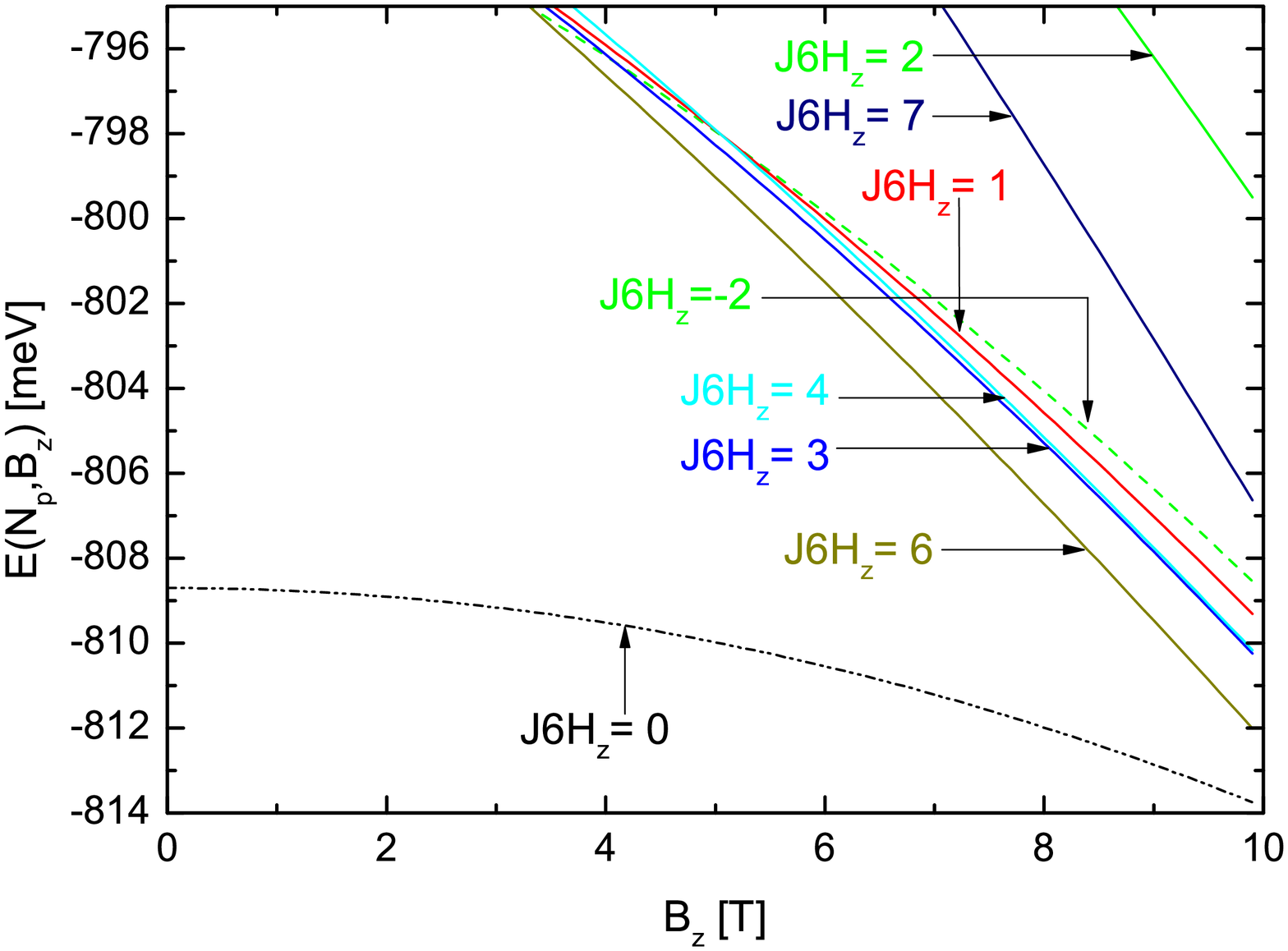}} &~~~&
\rotatebox{0}{\epsfxsize=70mm \epsfbox[65 35 705 505] {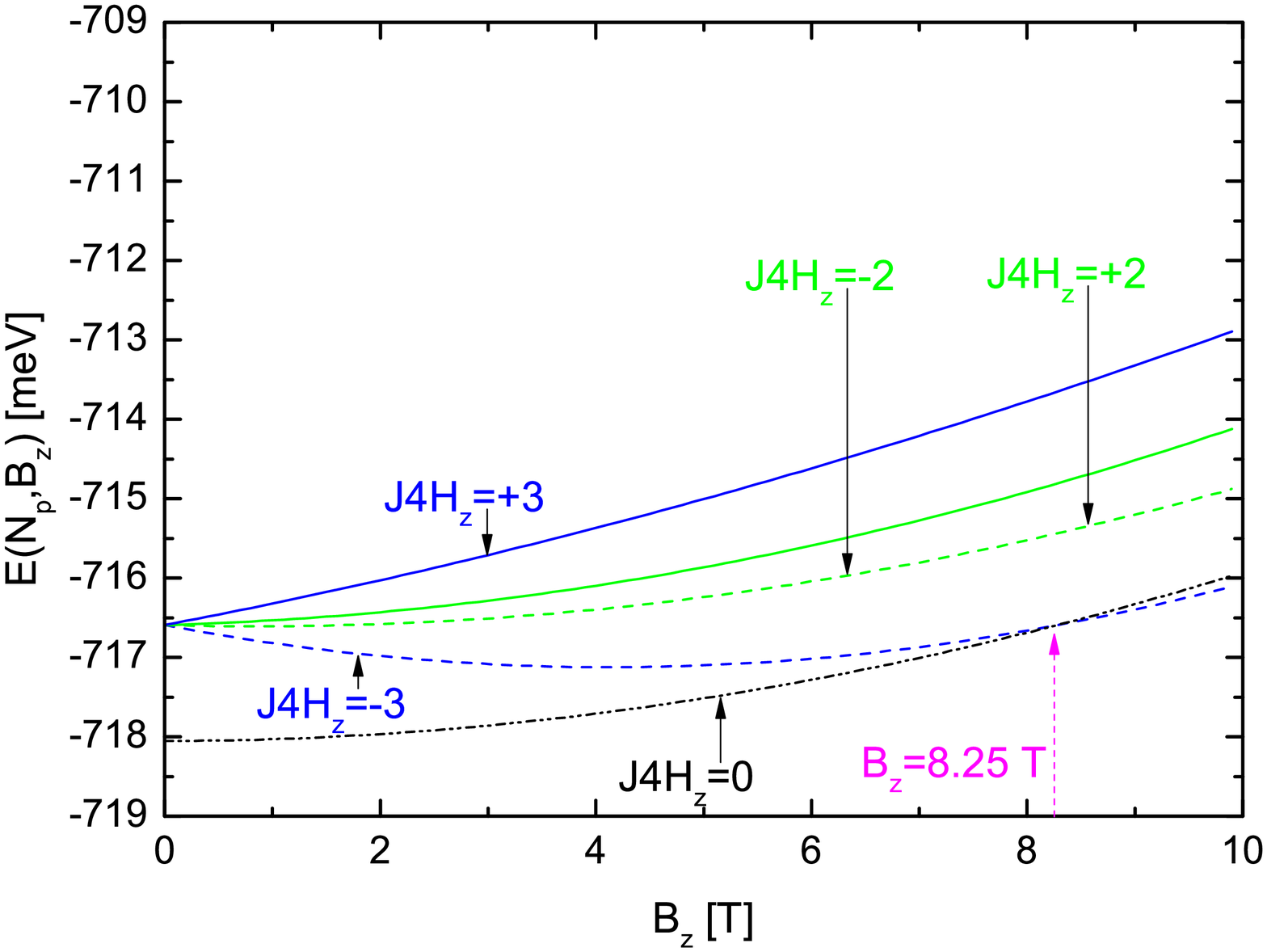}} \\
\end{tabular}}
\caption{(a-e) The energy spectra of multi-hole system in external magnetic field $B_z$; strong confinement. Eigenenergies for: (a) two holes, (b) three holes, (c) four holes, (d) five holes, (e) six holes. (f) The energy spectrum for four hole case with no Coulomb interaction included. Only states with $m_{J(N_P)H_z}=1$ are shown.\label{multihole-spectra}}
\end{figure*}

The energy spectra of two-hole to six-hole strongly confined systems are presented in Fig.~\ref{multihole-spectra}. As we are primarily interested in determining the ground state of the system, we only include the states with $m_{J(N_P)H_z}=1$ for each $J(N_P)H_z$, that is the lowest-energy level in each subspace defined by the multi-hole total angular momentum. The variational parameters of each computation are presented in Table \ref{TabVP}.

\begin{table}[ht]
\begin{center}
    $\begin{array}{c|c|c|c} \hline
        N_P  &  NV_A & NV_B & J(N_P)H_z \in\\ \hline
        2 & 100 & 100 & \{-9,...,9\}\\ \hline
        3 & 100 & 200 & \lbrace-15/2,...,15/2\rbrace\\ \hline
        4 & 100 & 300 & \{-8,...,8\}\\ \hline
        5 & 100 & 800 & \lbrace-17/2,...,17/2\rbrace\\ \hline
        6 & 100 & 1000 & \{-9,...,9\}\\ \hline                
    \end{array}$
\end{center}
\caption{Variational parameters for multi-hole systems. For the meaning of $N_P$, $NV_A$, $NV_B$ and $J(N_P)H_z$ symbols -- see the text.}\label{TabVP}\end{table}

If two holes are confined in the dot, the state with $J2H_z=0$ is the ground state as shown in Fig.~\ref{multihole-spectra}(a). There is not any state with different $J2H_z$ within close vicinity of this state on energy scale. Level next in order is the one corresponding to $J2H_z=-2$, separated by $18.6$ meV for $B_z=0$. In the case of three holes levels that correspond to states with $|J3H_z|=\frac{1}{2}$ and $|J3H_z|=\frac{5}{2}$ are close in terms of energy [see Fig.~\ref{multihole-spectra}(b)] -- the difference being equal to $0.9$ meV in the absence of magnetic field. For $B_z < 7.2$ T the $J3H_z=-\frac{1}{2}$ is the ground state but as the intensity of the field rises, the $J3H_z=-\frac{5}{2}$ level approaches the $J3H_z=-\frac{1}{2}$ one and eventually \mbox{above $7.2$ T} it becomes the ground state. When one more hole is added to quantum dot [$N_P=4$; Fig.~\ref{multihole-spectra}(c)] then for $B_z = 0$ the three states have the same energy: $J4H_z=-3$, $J4H_z=0$ and $J4H_z=+3$. In the presence of non-zero magnetic field this degeneracy is lifted and the $J4H_z=-3$ state is the ground state. When the dot is charged with five particles [Fig.~\ref{multihole-spectra}(d)] and $B_z=0$, then $|J5H_z|=\frac{1}{2}$ levels are relatively close to $|J5H_z|=\frac{5}{2}$ ones, with energy only higher by $1.2$ meV . Although the eigenenergy of $J5H_z = -\frac{13}{2}$ and $J5H_z = -\frac{7}{2}$ states strongly decreases with increasing magnetic field and the $J5H_z = -\frac{5}{2}$ level energy increases, the latter one has lowest energy in the spectrum for as much \mbox{as $10$ T}. For six-holes the $J6H_z=0$ level energy is by far lowest ($18.3$ meV) in low magnetic field regime. Although many other levels approach it with increasing magnetic field it remains the ground state in whole presented range of magnetic spectrum -- as shown \mbox{in Fig.~\ref{multihole-spectra}(e)}.

To sum up, the actual sequence of $J(N_P)H_z$ values of ground states for $N_P{\in}\{2,...,6\}$ is $\left(0,-\frac{1}{2},-3,-\frac{5}{2},0\right)$. If one compares this sequence to one that was deduced using Aufbau principle from single-hole calculations, it may be noticed that the principle correctly predicts $J(N_P)H_z$ for all cases apart from the four-hole one.

Let us consider how the magnitudes of energy separations between the respective single-hole states influence the behaviour of multiple-hole system. Firstly, the energy separation between the $J_z=\frac{3}{2}$ and $J_z=-\frac{1}{2}$ is large [see Fig.~\ref{onehole}(a)]. Then it is easy to predict that the dot charged with two holes will adhere to Aufbau principle as one-particle energy is dominant factor. The \mbox{${\sim}25$ meV} energy difference is an effective barrier to occupying any other orbitals apart of $|J_z|=\frac{3}{2}$ ones. A similar case is encountered for six-hole system. The first six one-hole levels are strongly separated from the seventh and the next ones ($22$ meV for $B_z=0$). It allows to predict that in six-hole wavefunction one will be facing occupation of orbitals corresponding to the mentioned states. In both cases two large energy separations in single-hole spectrum also translate to large separations in $N_P=2$ and $N_P=6$ ones -- see Fig.~\ref{multihole-spectra}(a) and Fig.~\ref{multihole-spectra}(e), respectively. Conversely, the relative vicinity of $|J_z|=\frac{1}{2}$ and $|J_z|=\frac{5}{2}$ states in single-hole spectrum of Fig.~\ref{onehole}(a) opens the possibility of occupation of orbitals other than the lowest $N_P$ ones in the case of three, four and five holes. As said before this possibility actually realizes only for $N_P=3$ and $N_P=4$.

Let us now focus on the direct cause of the breaking of the Aufbau principle in these two cases -- that is the Coulomb interaction. For $N_P=3$ in the range where the violation occurs, the $J_z=-\frac{5}{2}$ orbital is preferred to the $J_z=-\frac{1}{2}$ one. This allows us to deduce that the Coulomb repulsion between $J_z=-\frac{5}{2}$ and $J_z=\pm\frac{3}{2}$ orbitals is weaker that the between $J_z=-\frac{1}{2}$ and $J_z=\pm\frac{3}{2}$ orbitals. For $N_P=4$ analogically $J_z=-\frac{5}{2}$ is preferred to the $J_z=\frac{1}{2}$ one in respect of repulsion between it and other occupied states: $J_z=\pm\frac{3}{2}$ and $J_z=-\frac{1}{2}$.

In order to directly show that the Coulomb interaction is responsible for violation of the Aufbau principle, the spectrum Fig.~\ref{multihole-spectra}(f) is included. This figure shows the results for a $N_P=4$ model with no particle-particle interaction. As expected, the $J4H_z=0$ is the ground state for magnetic field up to $8.25$ T, which is the same $B_z$ value as for the crossing of $J_z=\frac{1}{2}$ and $J_z=-\frac{5}{2}$ levels in Fig.~\ref{onehole}(a).

\begin{figure*}[!ht]
 \makebox[\textwidth][c]{
\begin{tabular}{lcl}
(a) &~~~~~~& (b)\\
\rotatebox{0}{\epsfxsize=70mm \epsfbox[65 35 705 505] {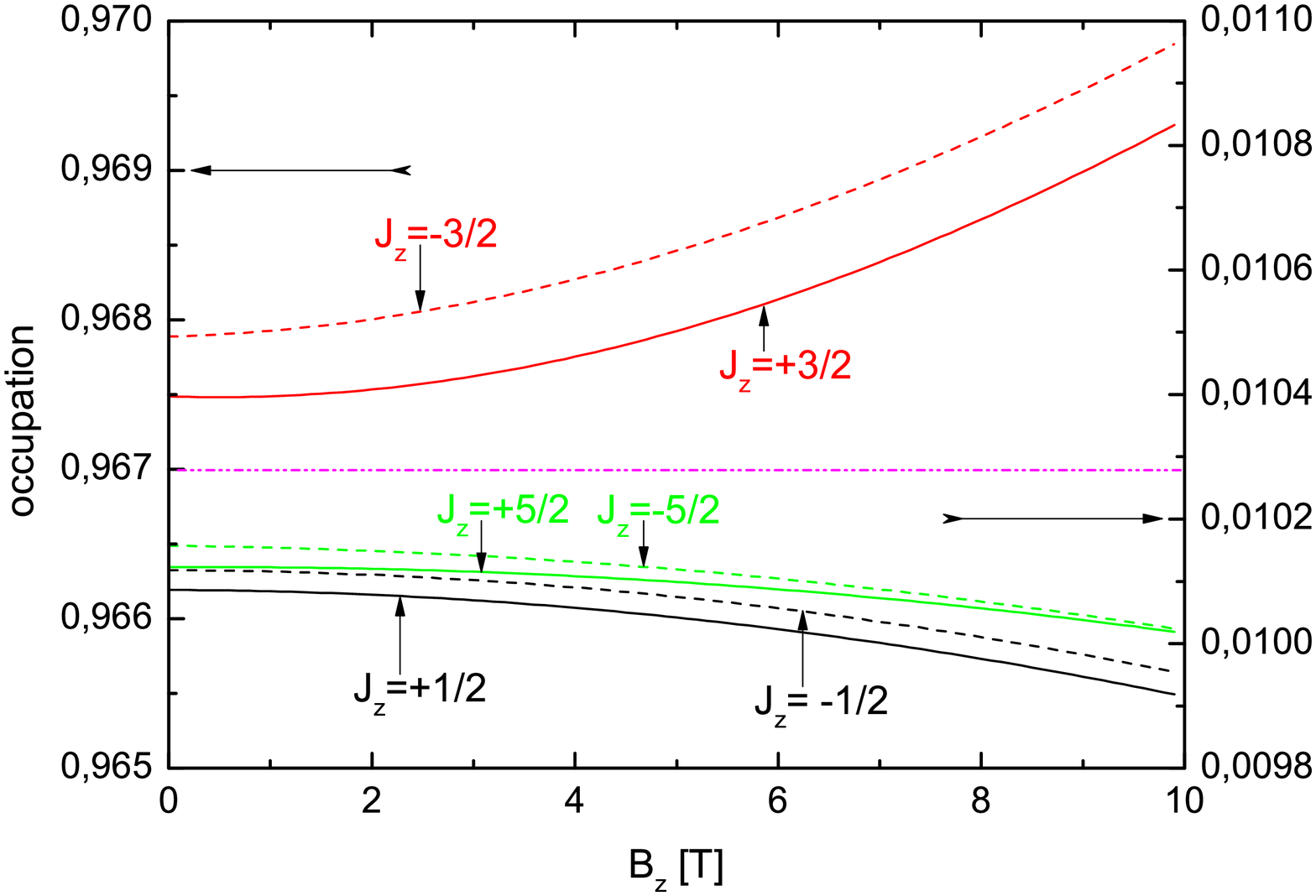}} &~~~~~~&
\rotatebox{0}{\epsfxsize=70mm \epsfbox[65 35 705 505] {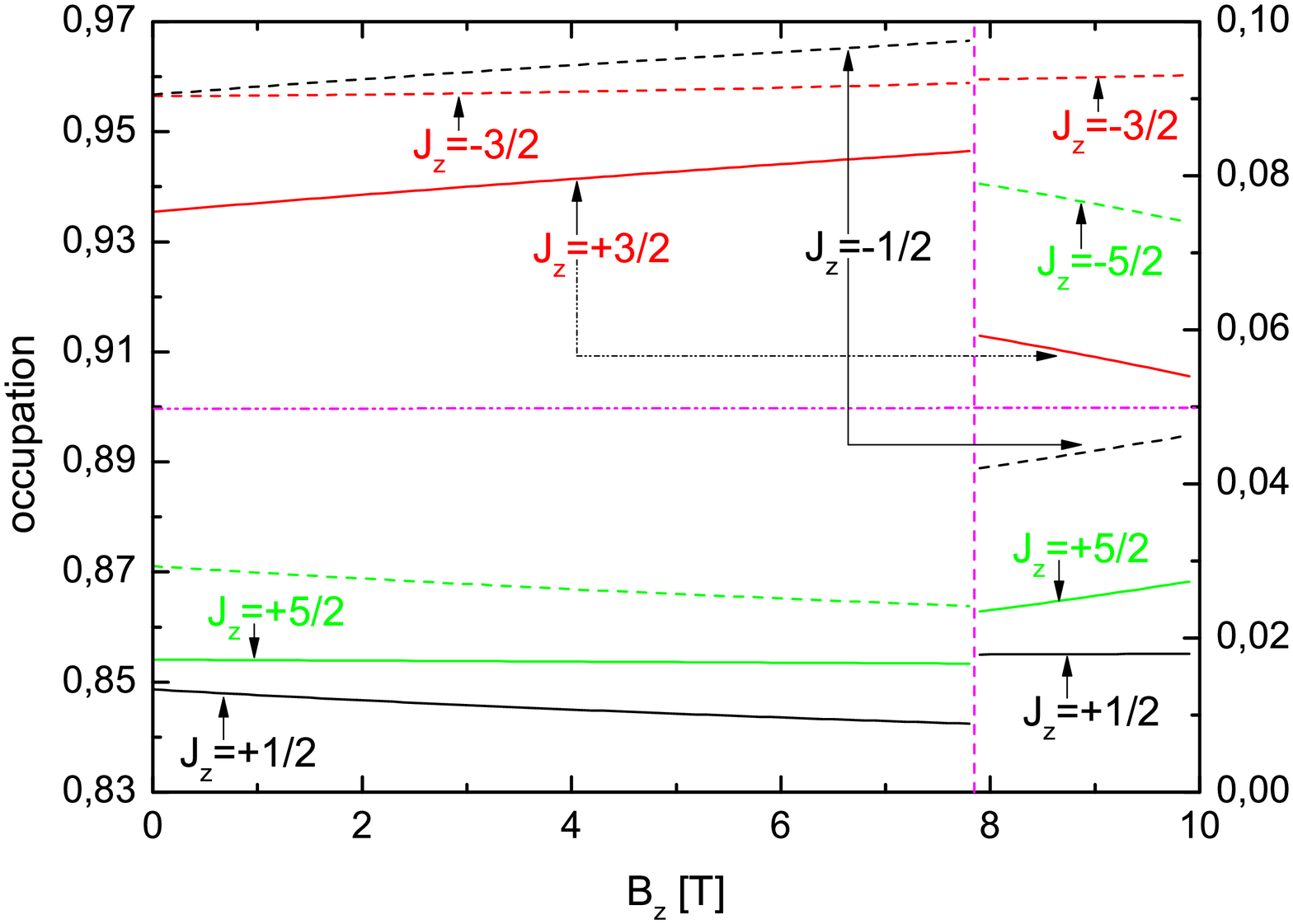}} \\
(c) &~~~~~~& (d)\\
\rotatebox{0}{\epsfxsize=70mm \epsfbox[65 35 705 505] {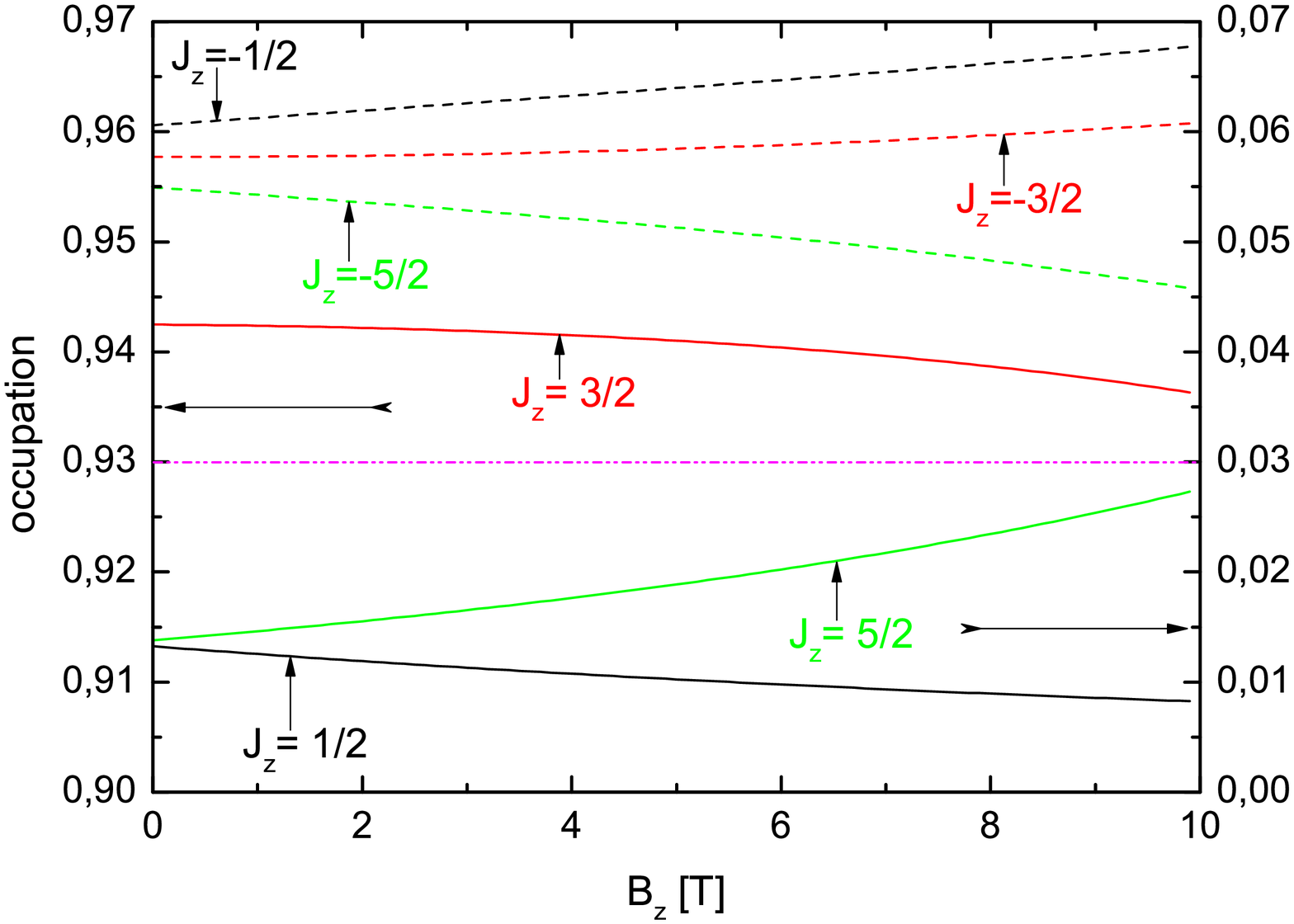}} &~~~~~~&
\rotatebox{0}{\epsfxsize=70mm \epsfbox[65 35 705 505] {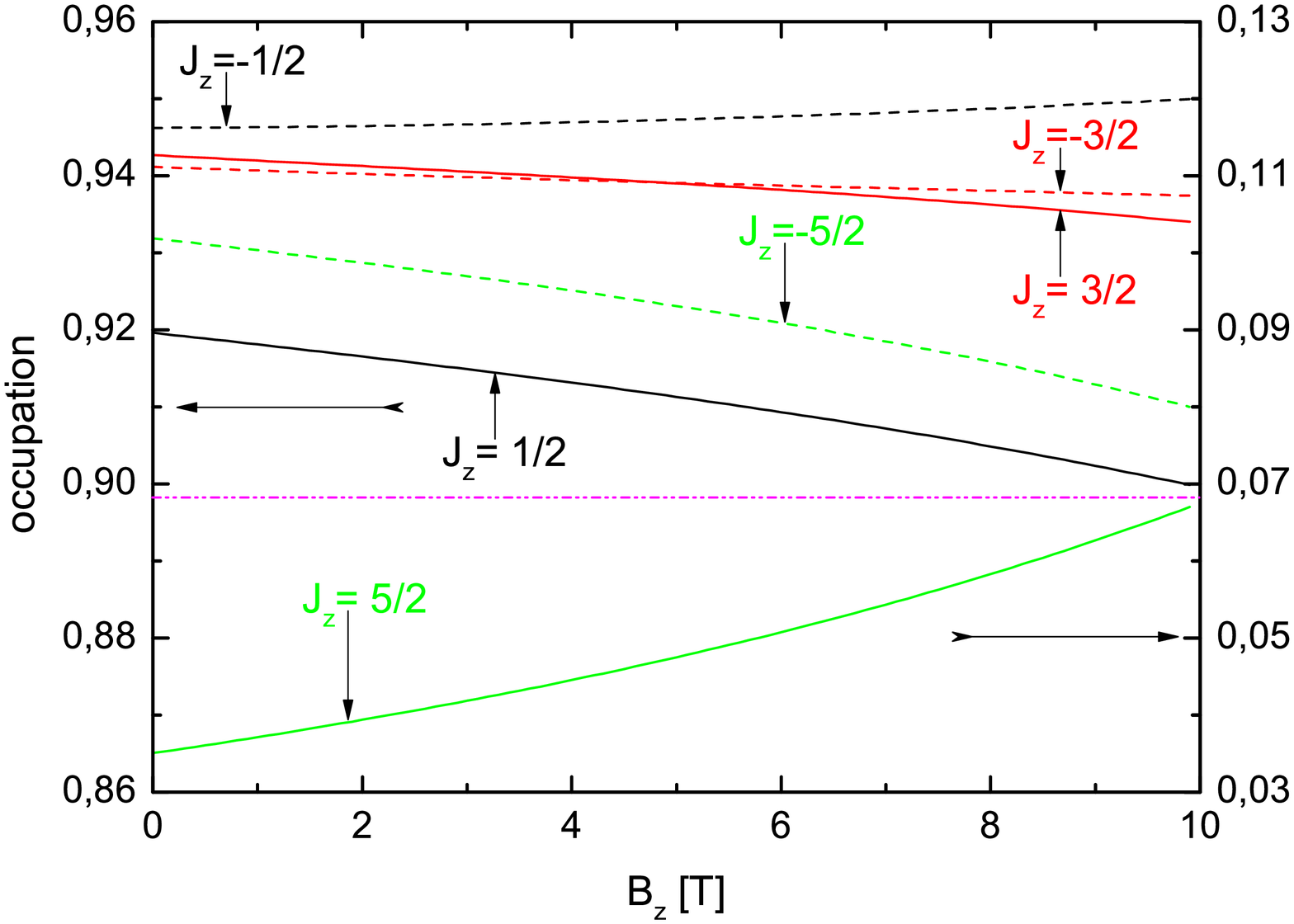}} \\
(e)\\
\rotatebox{0}{\epsfxsize=70mm \epsfbox[65 35 705 505] {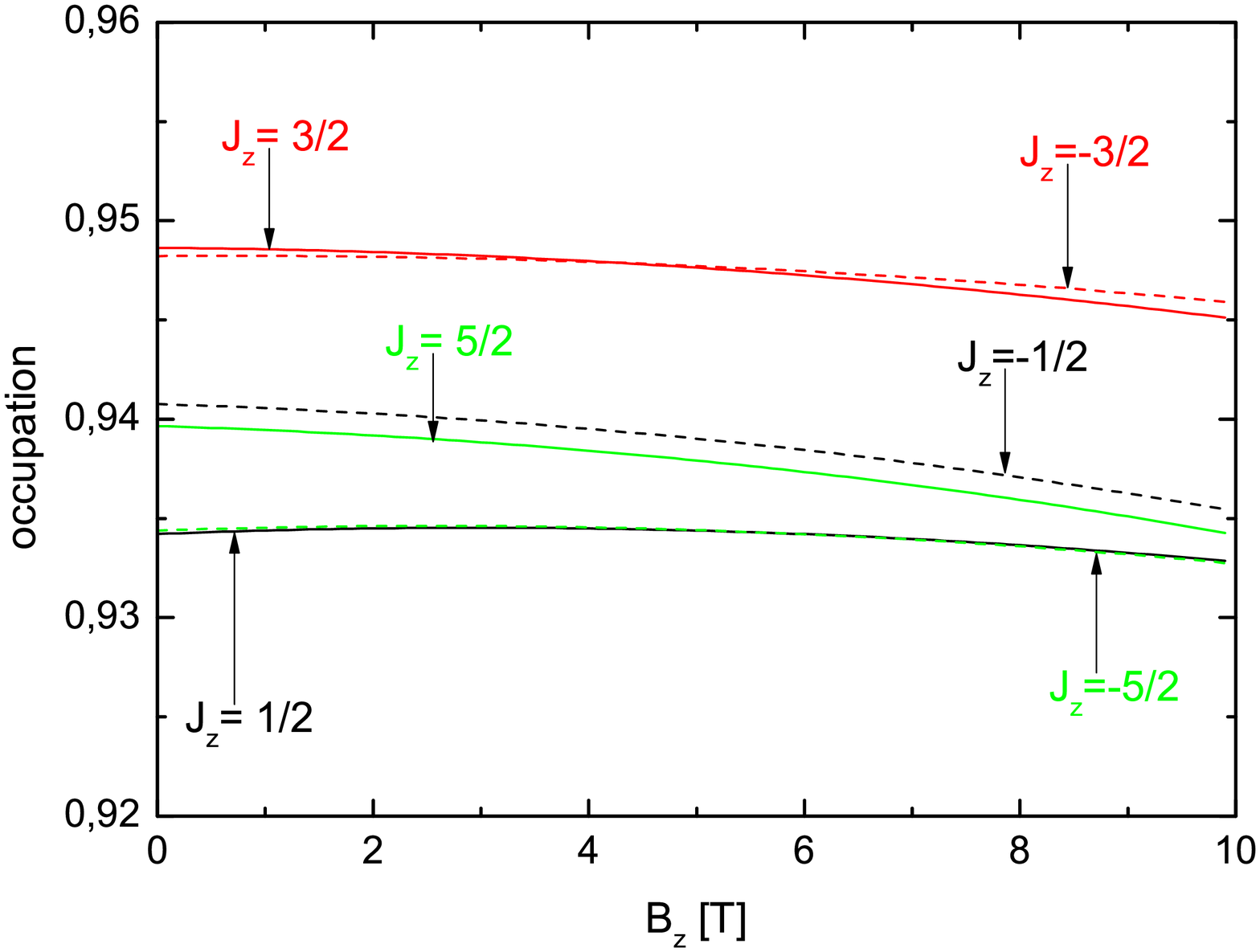}}
\end{tabular}}
\caption{(a-e) The occupation of first six single-hole orbitals in multi-hole system in external magnetic field $B_z$; strong confinement. The~case~of: (a) two holes, (b) three holes, (c) four holes, (d) five holes, (e) six holes. In (a-d) the horizontal dash-dot-dot line demarks the parts corresponding to right scale (bottom part) from the one corresponding to the left scale (top part). \label{wektory}}
\end{figure*}

An in-detail study of the system in regard to the Aufbau principle is enabled by obtaining the occupation coefficients of the six lowest-in-energy orbitals for the multi-hole ground states. As the multi-hole basis is constructed for no magnetic field, and it is taken into account at the same stage as the Coulomb interaction it is much more practical to use one-particle orbitals at $B_z=0$ for obtaining these coefficients instead of the actual orbitals at given $B_z>0$ . This is justified as i) we want only to discern the occupied orbitals from the unoccupied ones, and will not draw any conclusions from the details of the dependence of the coefficients on $B_z$ ii) the single-hole states may only mix in scope of the same $J_z$ subspace but for the whole concerned range of magnetic field the $m_{J_z}=1$ states are energetically separated from $m_{J_z}\geq2$ ones for each $J_z\in\lbrace\pm\frac{1}{2},\pm\frac{3}{2},\pm\frac{5}{2}\rbrace$ -- please note that this situation will differ for the weak confinement case.

The results for two holes are presented in Fig. \ref{wektory}(a). As expected, orbitals corresponding to the ground ($J_z=-3/2$) and first excited states ($J_z=3/2$) of the single hole are occupied. The relevant coefficients are equal to about $0.97$ for whole range of magnetic field. The other orbitals are empty, with occupation coefficients of circa $0.01$. The results for $N_P=3$ are presented in Fig. \ref{wektory}(b). In the low magnetic field range, where  ($J3H_z=-1/2$, $m_{-1/2}=1$) is the ground state of the system, the first three orbitals are occupied: $\lbrace J_z={\pm}3/2, J_z=-1/2 \rbrace$ and with relevant coefficients of $0.94$ or larger. Three next orbitals are not occupied: $\lbrace J_z=1/2, J_z={\pm}5/2 \rbrace$ with coefficients below $0.03$. At the point of the crossing ($B_z=7.81$ T) the character of the ground state changes to ($J3H_z=-5/2$, $m_{-5/2}=1$) and -- as expected -- an electron is transferred from occupying the $J_z=-\frac{1}{2}$ orbital to $J_z=-\frac{5}{2}$ one. At this point one should note that this crossing has no corresponding one in the single hole spectrum [see Fig.~\ref{onehole}(a)]. The three orbitals lowest-in-energy for any value of the magnetic field in the considered range are: ($J_z=-\frac{3}{2}, m_{-\frac{3}{2}}=1$), ($J_z=\frac{3}{2}, m_{\frac{3}{2}}=1$) and ($J_z=-\frac{1}{2}, m_{-\frac{1}{2}}=1$). The system follows the Aufbau principle $B_z < 7.81$ T  but the principle is clearly violated for large magnetic field. An another instance of its violation is the case of the $N_P=4$ -- as presented in Fig. \ref{wektory}(c). Although the $J_z={\pm}3/2$ orbitals are occupied, the $J_z={\pm}1/2$ are not paired. The $Jz=-5/2$ orbital is occupied instead of the $J_z=1/2$ one for the whole considered range of magnetic field. Please note that in Fig.~\ref{multihole-spectra}(c) there is not any signature of the $B_0$ crossing in Fig.~\ref{onehole}(a). This is an example of breaking of the Aufbau principle, even more clear than in the $N_P=3$ case because it happens for any $0<B_z<B_0$, so it obviously may not be interpreted as inducted by a strong magnetic field. Our system returns to previous behaviour, when an additional, fifth hole is confined in the dot. As shown in Fig. \ref{wektory}(d) the orbitals that correspond to the one-particle states of five lowest energies are occupied with their occupation coefficients of about $0.91$ to $0.95$, depending on orbital. The $J_z=5/2$ orbital have occupation coefficient lower than $0.07$. All six one-particle orbitals are occupied with occupation coefficients of over $0.93$ when six holes are present in the system, as shown in Fig. \ref{wektory}(e).

\subsubsection{Chemical potential}
\begin{figure*}[!ht]
 \makebox[\textwidth][c]{
\begin{tabular}{ll}
(a) & (b)\\
\rotatebox{0}{\epsfxsize=70mm \epsfbox[65 35 705 505] {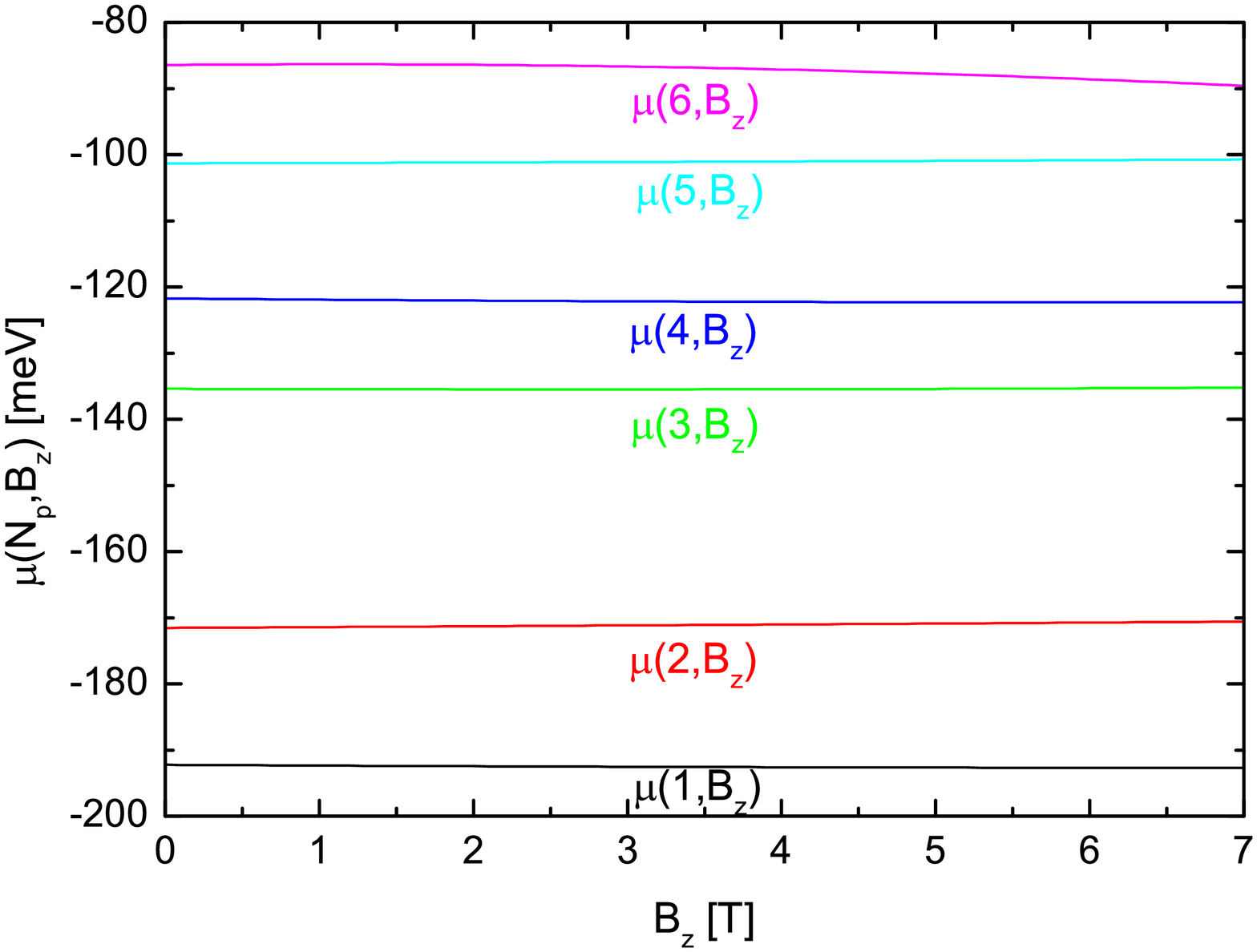}} &
\rotatebox{0}{\epsfxsize=70mm \epsfbox[65 35 705 505] {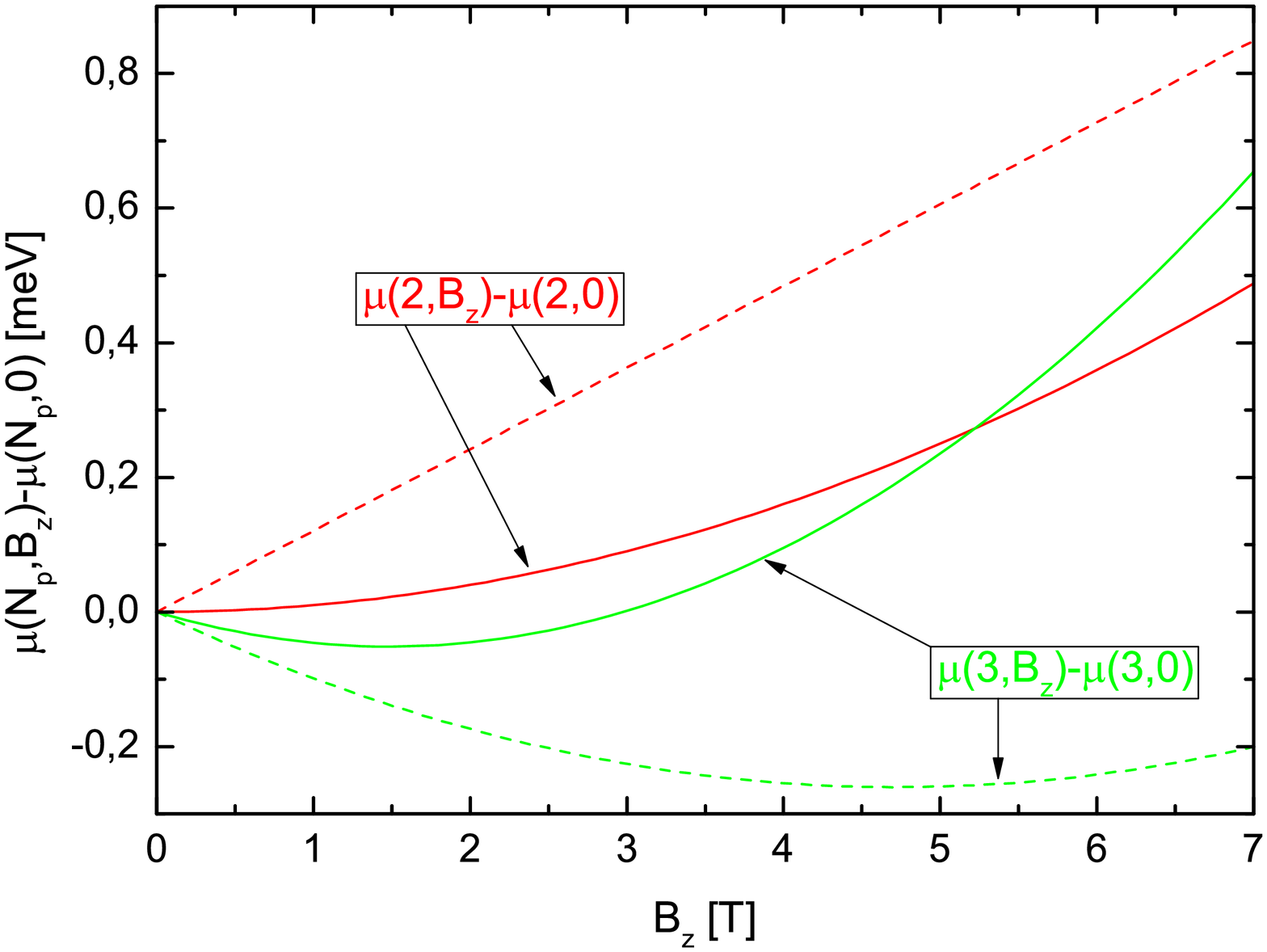}} \\
(c)\\
\rotatebox{0}{\epsfxsize=70mm \epsfbox[65 35 705 505] {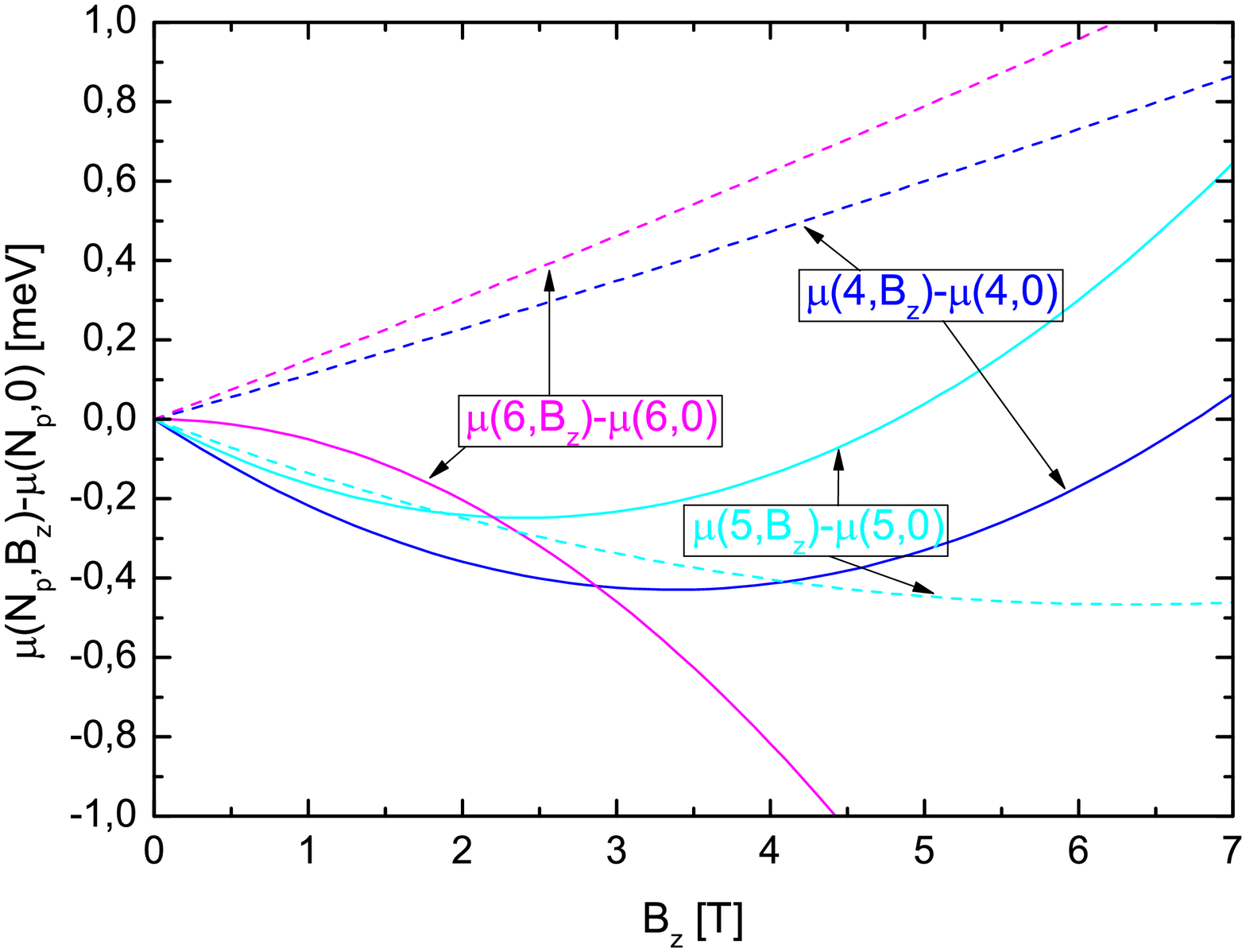}} \\
\end{tabular}}
\caption{(a) The chemical potential as a function of external magnetic field $B_z$ for $N_p{\in}\{1,...,6\}$; strong confinement. (b-c) Relative dependence of $\mu(N_p,B_z)$ on $B_z$ for: (b) $N_p{\in}\{2,3\}$, (c) $N_p{\in}\{4,5,6\}$. Solid lines correspond to the full model, and dashed lines to the non-interacting case.\label{chempot}}
\end{figure*}

The chemical potential of the system ${\mu}(N_p,B_z)=E(N_p,B_z)-E(N_p-1,B_z)$ is presented in Fig. \ref{chempot}(a). For up to five holes, the chemical potential is (in energy scale as presented in this figure) very weakly dependent on external magnetic field in the growth direction. For $N_p=6$ this dependence is much stronger, with the value of $\mu(N_p,B_z)$ decreasing with increasing $B_z$. Let us compare these results with the non-interacting case. When there is no hole-hole Coulomb interaction then $\mu(N_p,B_z)$ is equal to the energy of $N_p$-th single-hole state and hence the chemical potential spectrum is the same as the single-hole energy spectrum -- Fig. \ref{onehole}(a). The chemical potential for $N_p=1$ is trivially the same in both models. For $N_p>1$ the values of $\mu(N_p,B_z)$ of both models are completely different; compare Fig. \ref{chempot}(a) and Fig. \ref{onehole}(a). This difference is quite obvious as the non-interacting model does not account for upward shift in energies/chemical potential due to repulsive character of the electrostatic hole interaction. 

In order to compare the \textit{relative} dependence of chemical potential on magnetic field in both models we calculated $\mu(Np,Bz)-\mu(Np,0)$. The results are shown in Fig \ref{chempot}(a,b) and as it can be seen, also respective \textit{relative} dependence functions are completely different in the case of the normal computation (solid lines) and the non-interacting case (dashed lines). For example, while $\mu(2,Bz)-\mu(2,0)$ in the case of the simplified approach is linear-like function, it is clearly quadratical for the full model. Furthermore, the sign of $\mu(6,Bz)-\mu(6,0)$ is opposite in one model with respect to the other one. In conclusion, it is not possible to infer directly the behaviour of the chemical potential in magnetic field $B_z$ from the single-hole spectrum of the dot.

\subsection{Weak confinement}

In this section we present the results for the weak confinement case \textit{i.e.} the system with $R_{dot}=20$ nm and $2 Z_{dot}=6$ nm. The volume of this dot is twelve times bigger than the volume of the former one.

\subsubsection{Single hole}

\begin{figure*}[!ht]
\makebox[\textwidth][c]{
\begin{tabular}{lcl}
(a) &~~~~~~& (b)\\
\rotatebox{0}{\epsfxsize=65mm \epsfbox[65 35 705 505] {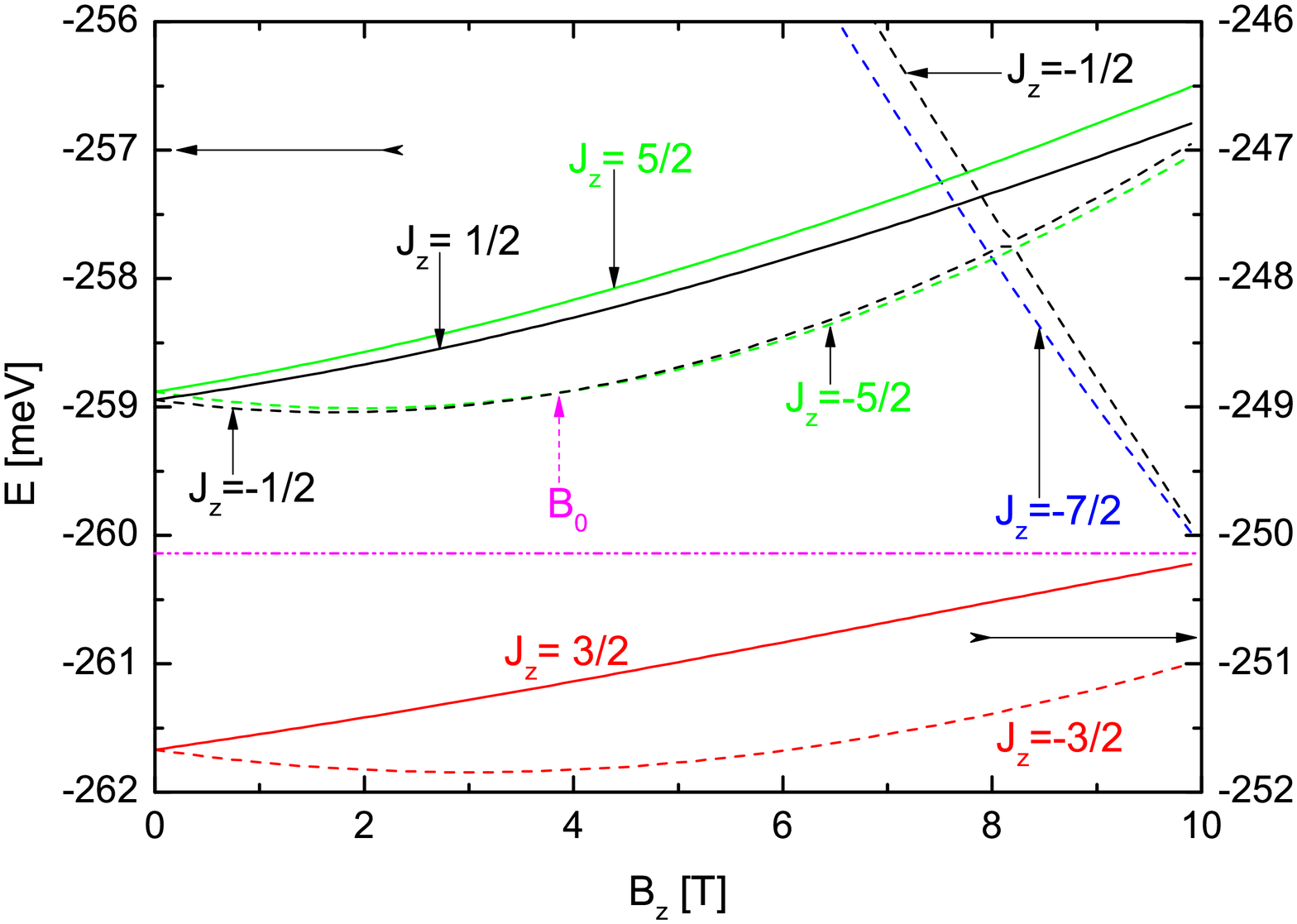}} &~~~~~~&
\rotatebox{0}{\epsfxsize=65mm \epsfbox[65 35 705 505] {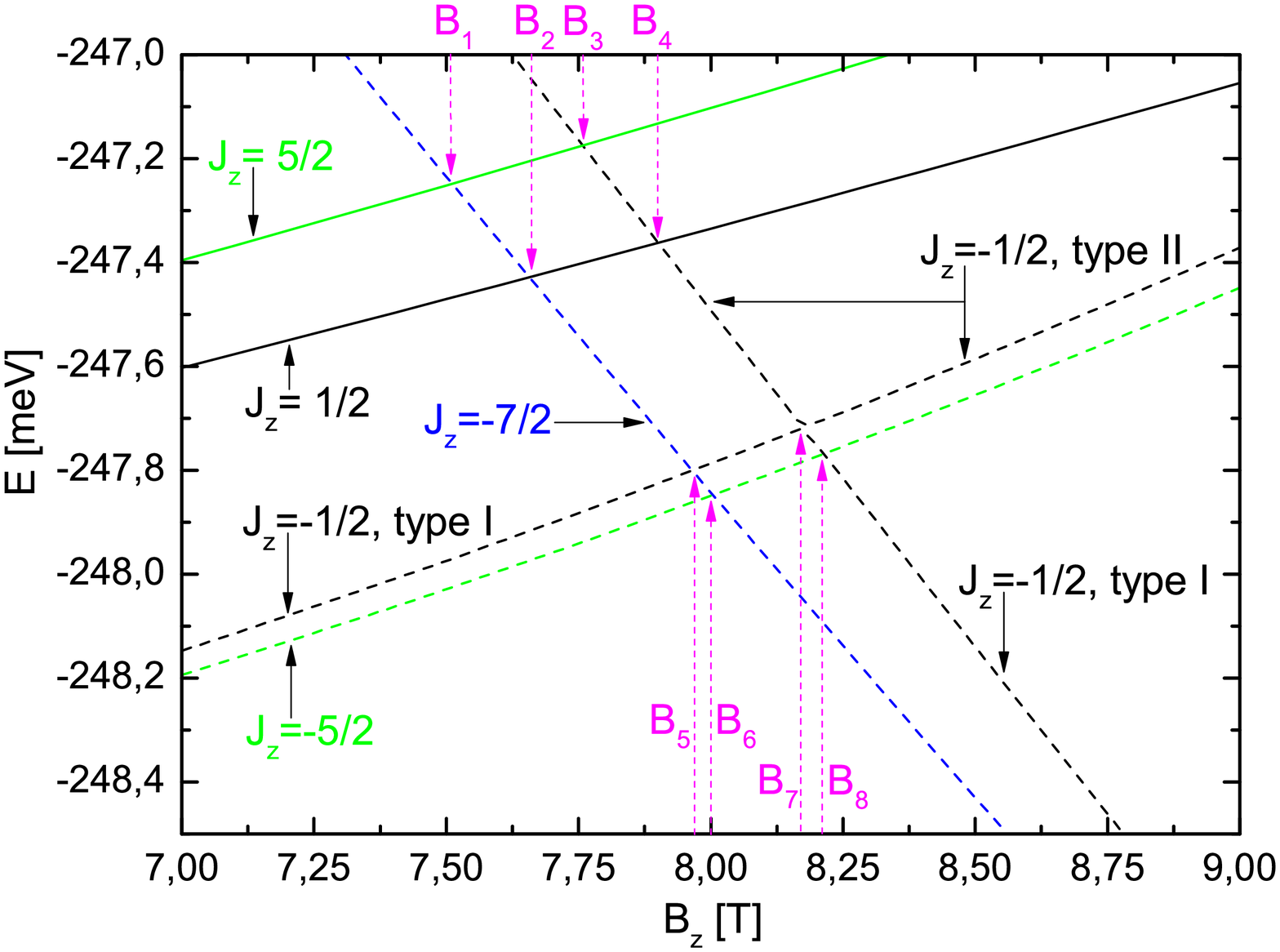}} \\
\end{tabular}}
\caption{(a) The energy spectrum of one hole in external magnetic field $B_z$; weak confinement case. (b) Close-up of (a) showing the region where a series of level crossings take place. For the details on the level crossings, see Table \ref{TabCrossings}.\label{WC-onehole}}
\end{figure*}

The energy spectrum of a single hole in external magnetic field $B_z$ for the weak confinement case is presented if Fig.~\ref{WC-onehole}(a). For low magnetic field the two energy levels of lowest energy are: ($J_z=-\frac{3}{2}, m_{-\frac{3}{2}}=1$), ($J_z=\frac{3}{2}, m_{\frac{3}{2}}=1$). The next four levels, \textit{i.e.} ($J_z=-\frac{1}{2}, m_{-\frac{1}{2}}=1$), ($J_z=\frac{1}{2}, m_{\frac{1}{2}}=1$), ($J_z=-\frac{5}{2}, m_{-\frac{5}{2}}=1$), ($J_z=\frac{5}{2}, m_{\frac{5}{2}}=1$) are nearly degenerate at $B_z\sim0$. This is in agreement with a simple intuition that links larger dimensions of the quantum dot with smaller energy differences of quantized levels. As the intensity of the field increases a pair of states with negative $J_z$ becomes significantly separated from the positive $J_z$ pair. However, the energy separation as the difference between $J_z=\frac{1}{2}$ and $J_z=\frac{5}{2}$ states (that will be called \textit{higher-energy pair}) and -- especially -- the difference between the $J_z=-\frac{1}{2}$ and $J_z=-\frac{5}{2}$ (that will be called \textit{lower-energy pair}) remains relatively small. Although the latter two levels are indeed slightly separated and their order changes in crossing at $B_0$ (see Table~\ref{TabCrossings}), they are nearly degenerated.

For a magnetic field interval of $B_z{\in}(7,9)$ T there is a series of level crossings that change the character of the levels from third to eighth, respectively. This part of the spectrum is presented in more detail in Fig.~\ref{WC-onehole}(b). The details concerning each crossing can be found in Table~\ref{TabCrossings}. 

\begin{table}[ht]
\begin{center}
    \begin{tabular}{C{25mm}|C{25mm}|C{25mm}} \hline
        $J_z$-s of engaged levels & ordinals of engaged levels & $B_z$ at crossing\\ \hline
        $-{5/2}$, $-{1/2}$ & $3,4$ & $B_0=3.86$ T \\ \hline
        $-{7/2}$, ${5/2}$ & $6,7$ & $B_1=7.51$ T \\ \hline
        $-{7/2}$, ${1/2}$ & $5,6$ & $B_2=7.66$ T \\ \hline
        $-{1/2}$, ${5/2}$ & $7,8$ & $B_3=7.76$ T \\ \hline
        $-{1/2}$, ${1/2}$ & $6,7$ & $B_4=7.90$ T \\ \hline
        $-{7/2}$, $-{1/2}$ & $4,5$ & $B_5=7.97$ T \\ \hline
        $-{7/2}$, $-{5/2}$ & $3,4$ & $B_6=8.00$ T \\ \hline
        $-{1/2}$, type I/II & $5,6$ & $B_7=8.17$ T \\ \hline
        $-{5/2}$, $-{1/2}$ & $4,5$ & $B_8=8.21$ T \\ \hline 
    \end{tabular}
\end{center}
\caption{The details concerning level crossings of~Fig.~\ref{WC-onehole}(a,b)}\label{TabCrossings}\end{table}

Here a short digression is necessary. One should notice that in the case of the weak confinement in the single-hole spectrum there is a crossing of the states with the same $J_z=-\frac{1}{2}$. We want to analyse the occupation coefficients for multi-hole states in the same way, as it was done for the system with strong confinement, thus for this purpose these two orbitals will be referred to by their orbital character at $B_z=0$. The orbital that corresponds to $J_z=-\frac{1}{2}, m_{-\frac{1}{2}}=1$ level \underline{at $B_z=0$} will be refered to as \textit{type I}, and the one that corresponds to $J_z=-\frac{1}{2}, m_{-\frac{1}{2}}=2$ level \underline{at $B_z=0$} -- as \textit{type II}. Appropriate marks were put on Fig.~\ref{WC-onehole}(b) and in Table~\ref{TabCrossings}.

At this point one should note that the fact of $J_z=-\frac{1}{2}$ and $J_z=-\frac{5}{2}$ states being in a near-degeneracy for a significant part of the spectrum is a remarkable difference between the strong confinement case [Fig.~\ref{onehole}(a)] and this one. The consequence of this situation is that in the non-interacting picture for $N_P=3$ the energy separation of ground and first excited states will be very small. This allows to suspect that the characteristics of the actual relevant multi-hole spectrum will be very essentially dependent on the Coulomb interaction between the holes. Furthermore, the presence of a series of level crossings in the upper right part of the single-hole case suggests that some signatures of it may be found in multi-hole spectra.

\subsubsection{Energy spectra of multiple holes}

\begin{figure*}[!ht]
 \makebox[\textwidth][c]{
\begin{tabular}{lcl}
(a) &~~~~& (b)\\
\rotatebox{0}{\epsfxsize=70mm \epsfbox[65 35 705 505] {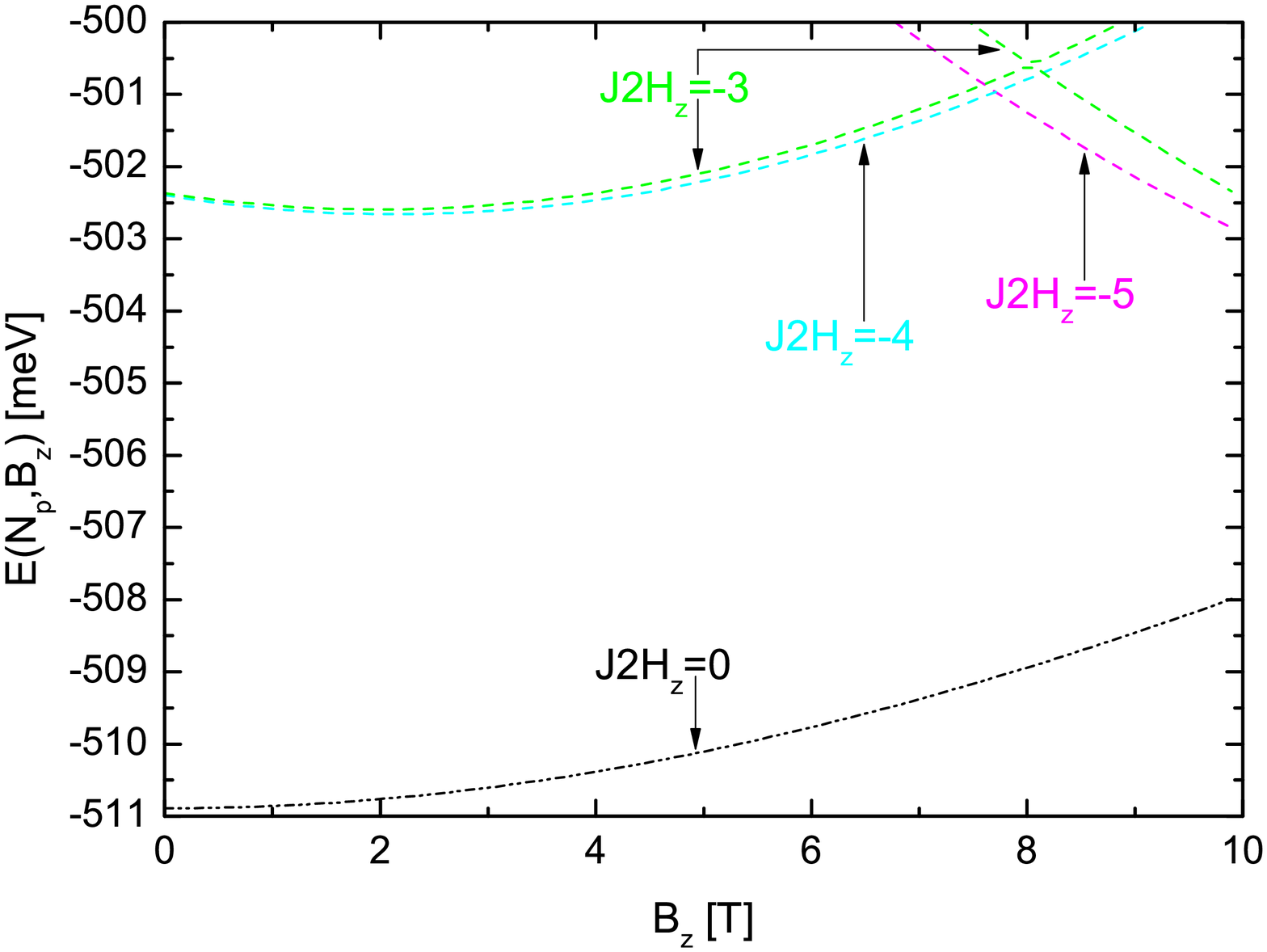}} &~~~~&
\rotatebox{0}{\epsfxsize=70mm \epsfbox[65 35 705 505] {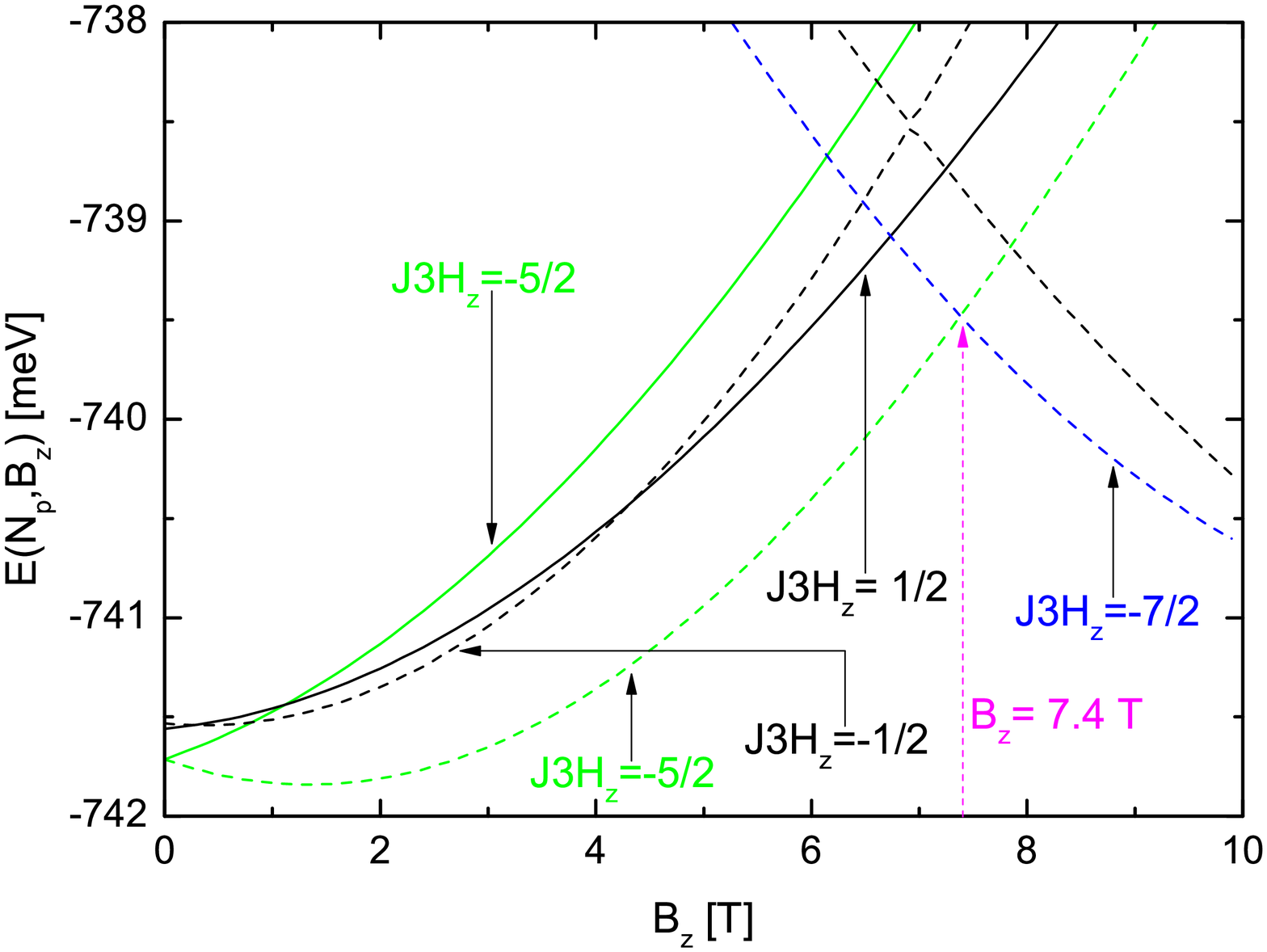}} \\
(c) &~~~~& (d)\\
\rotatebox{0}{\epsfxsize=70mm \epsfbox[65 35 705 505] {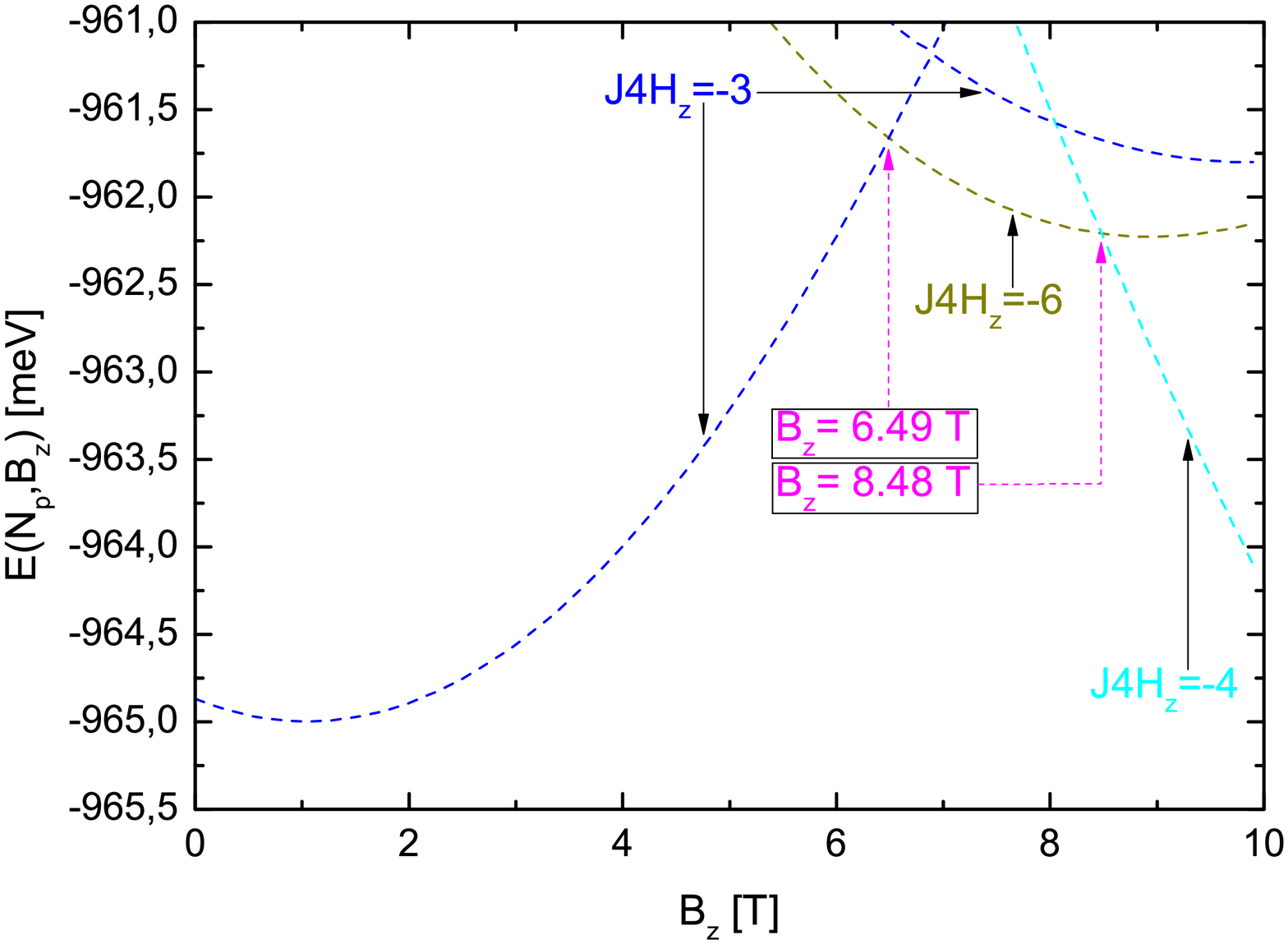}} &~~~~&
\rotatebox{0}{\epsfxsize=70mm \epsfbox[65 35 705 505] {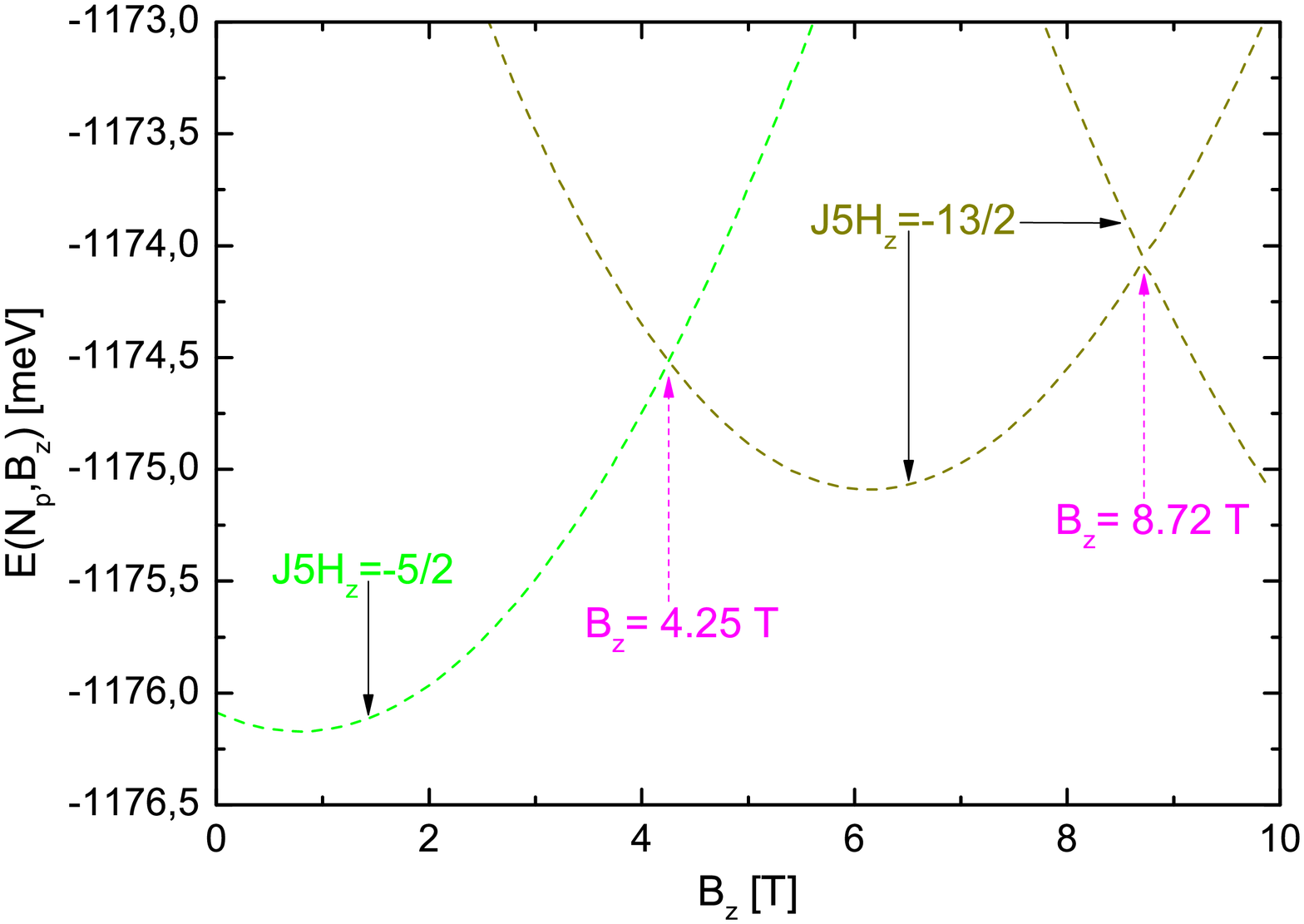}} \\
(e)\\
\rotatebox{0}{\epsfxsize=70mm \epsfbox[65 35 705 505] {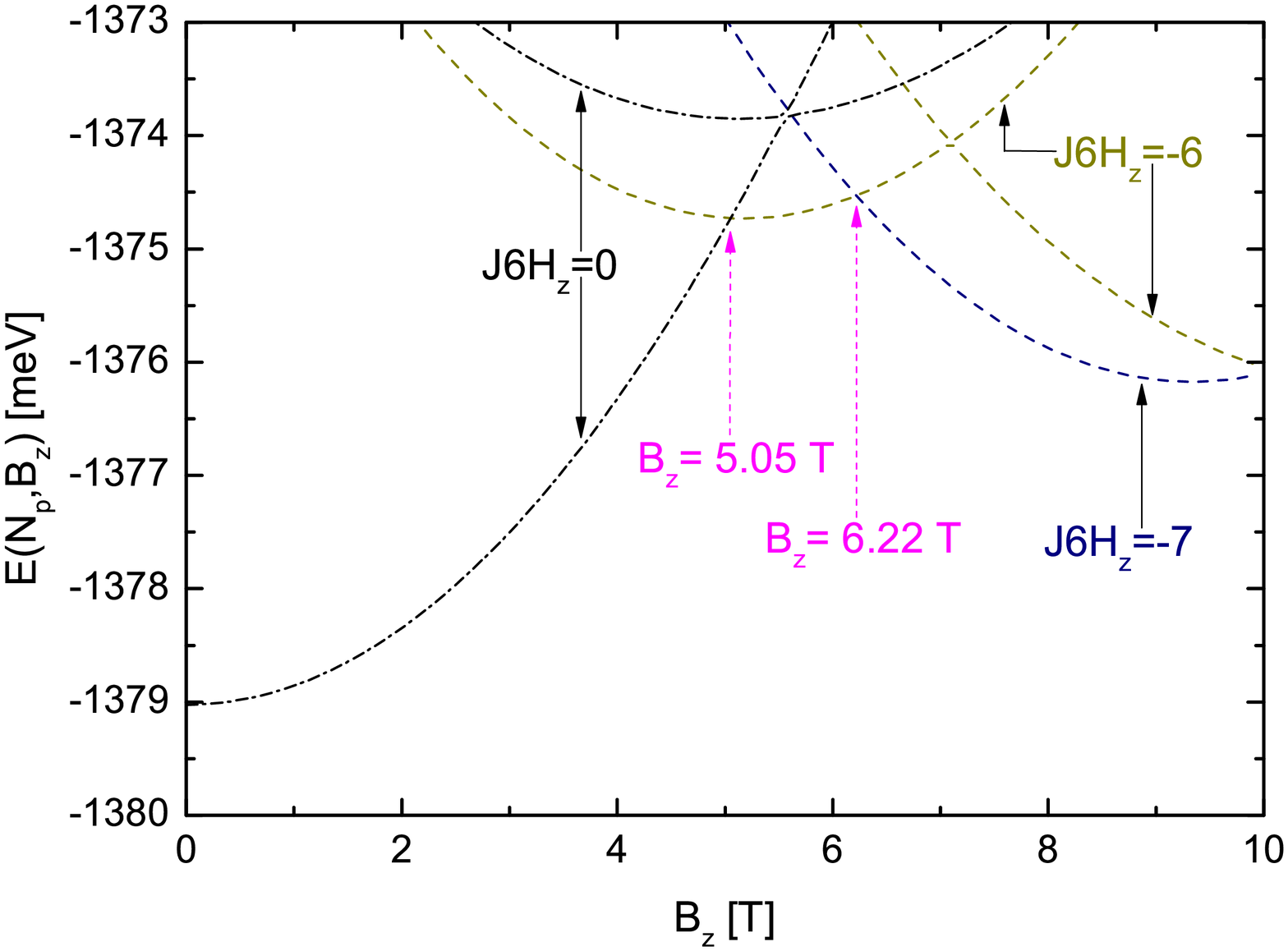}} \\
\end{tabular}}
\caption{(a-e) The energy spectra of multi-hole system in external magnetic field $B_z$; weak confinement case. Eigenenergies~for: (a) two holes, (b) three holes, (c) four holes, (d) five holes, (e) six holes. For $N_P\geq4$ only states that are relevant for determining the character and energy of the ground state level are shown.\label{WC-multihole-spectra}}
\end{figure*}

The energy spectra of two-hole to six-hole weakly confined systems are presented in Fig.~\ref{WC-multihole-spectra}. The bigger-dot case means there is a multitude of excited states in the multiple-hole spectra and including them all would render figures unreadable. Hence, for $N_P{\geq}4$, only a "relevant" subset of levels is shown, which includes the ground state and all levels that interact with it in any way. The computation was conducted with variational parameters as presented in Table \ref{TabVP}.

\begin{figure*}[!ht]
 \makebox[\textwidth][c]{
\begin{tabular}{lcl}
(a) &~~~~& (b)\\
\rotatebox{0}{\epsfxsize=70mm \epsfbox[65 35 705 505] {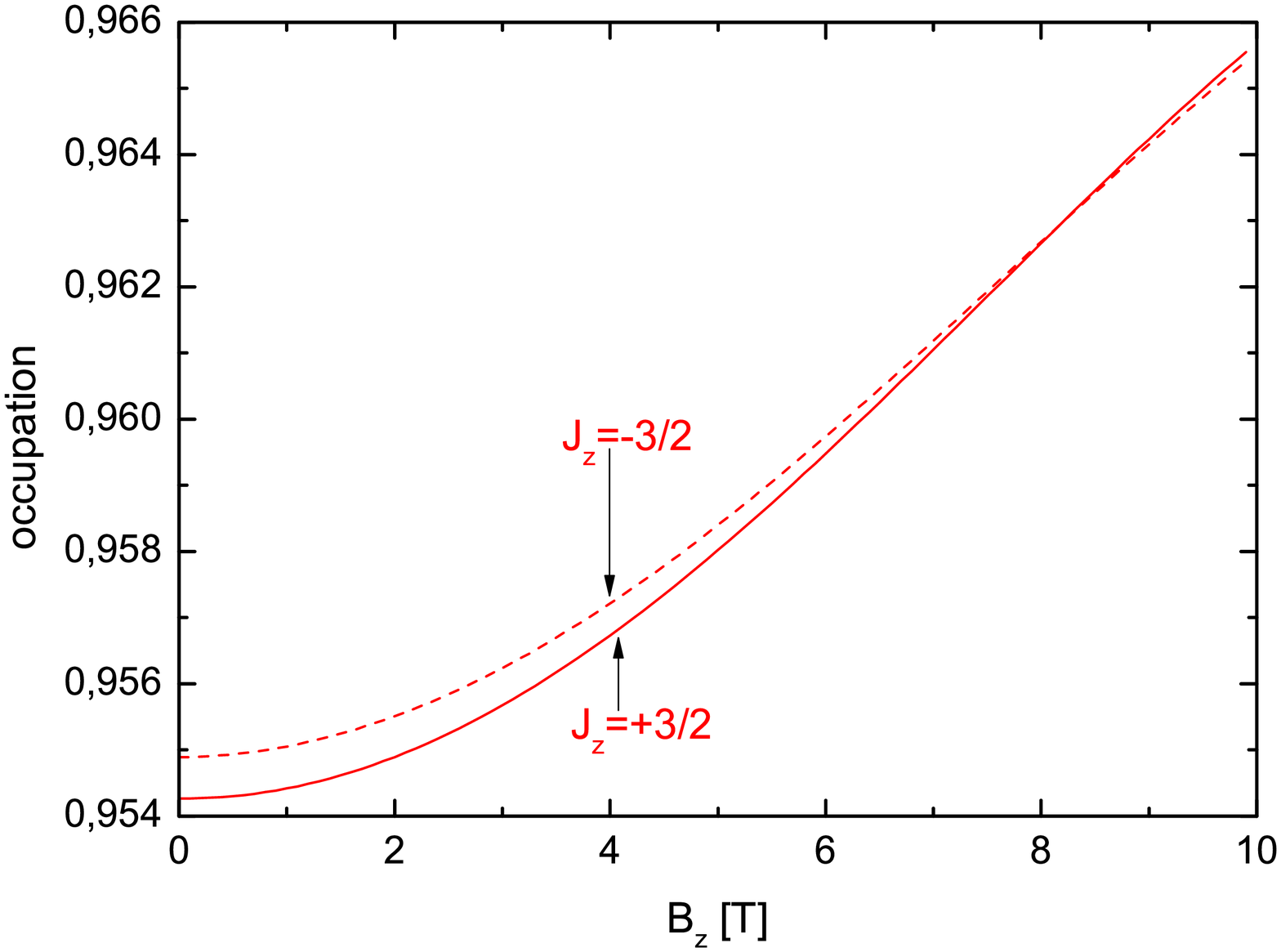}} &~~~~&
\rotatebox{0}{\epsfxsize=70mm \epsfbox[65 35 705 505] {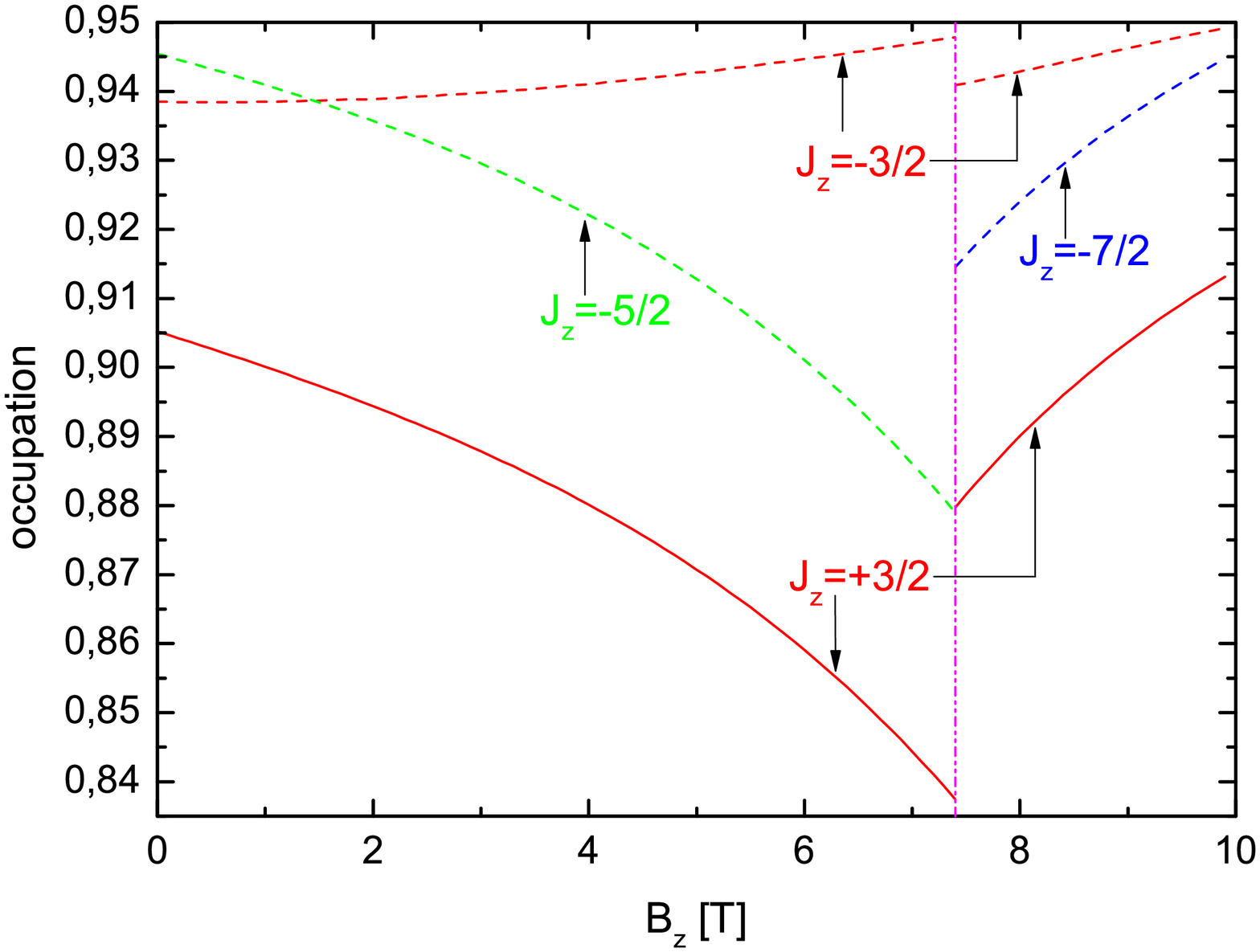}} \\
(c) & & (d)\\
\rotatebox{0}{\epsfxsize=70mm \epsfbox[65 35 705 505] {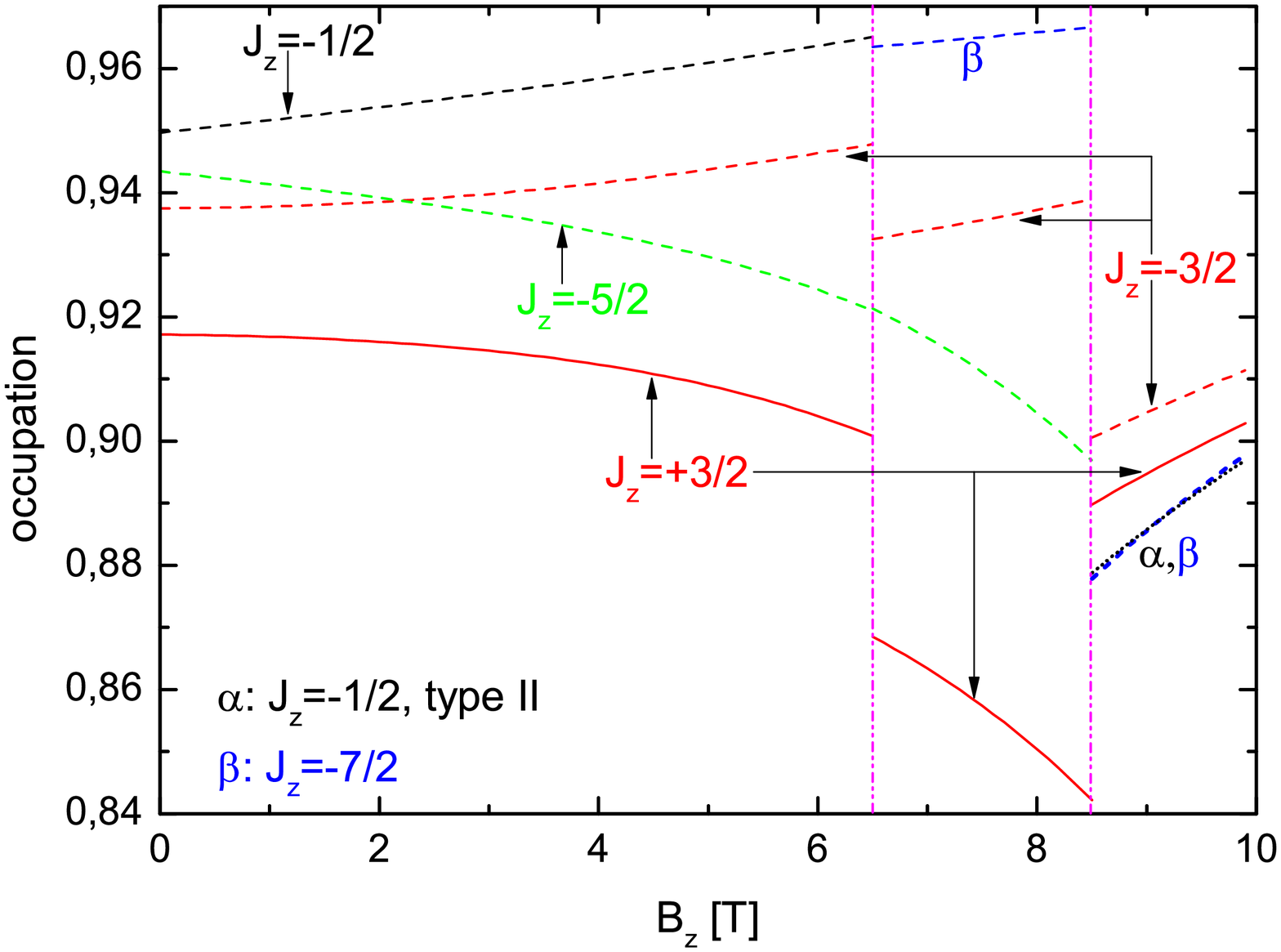}} &~~~~&
\rotatebox{0}{\epsfxsize=70mm \epsfbox[65 35 705 505] {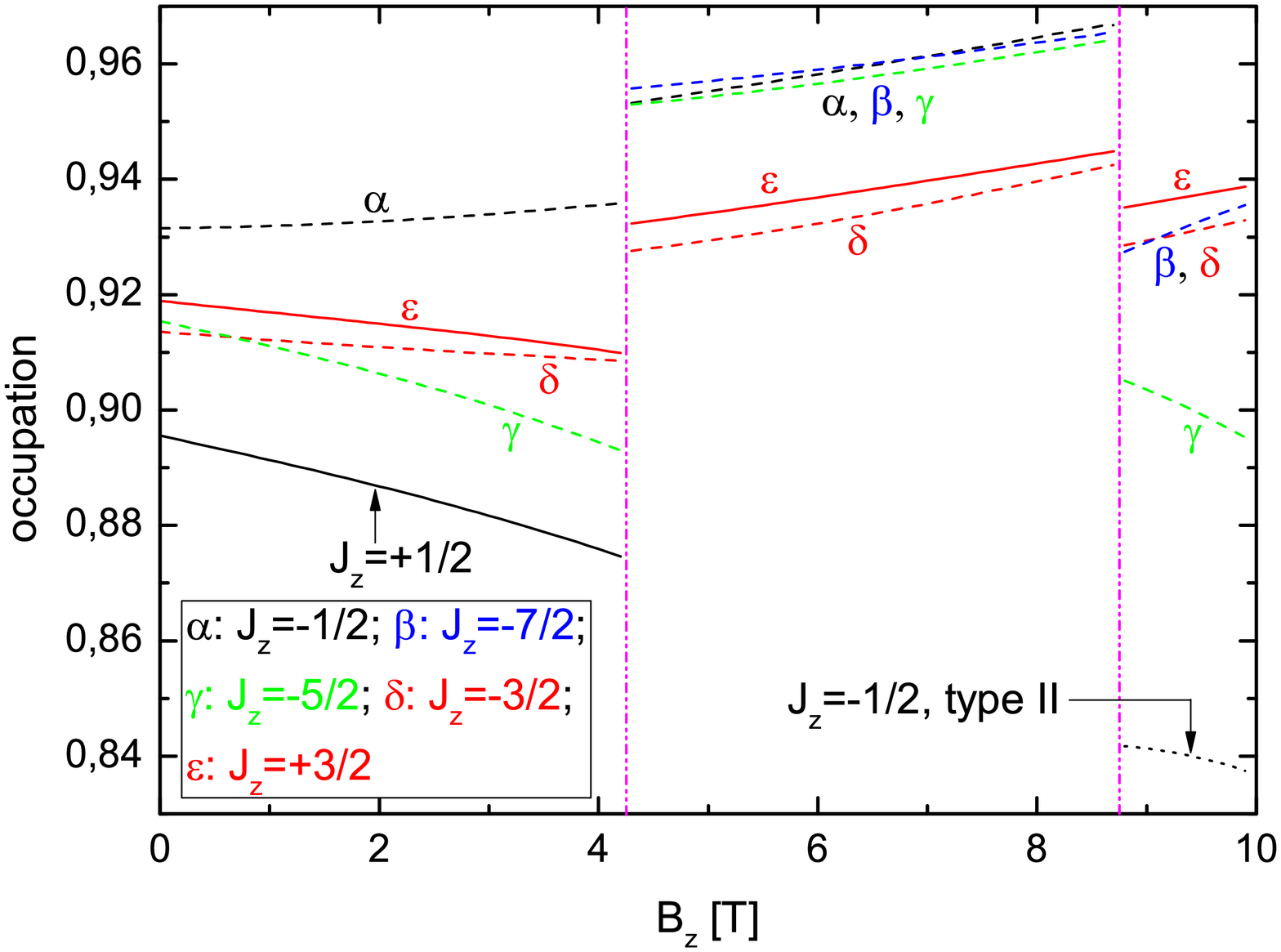}} \\
(e)\\
\rotatebox{0}{\epsfxsize=70mm \epsfbox[65 35 705 505] {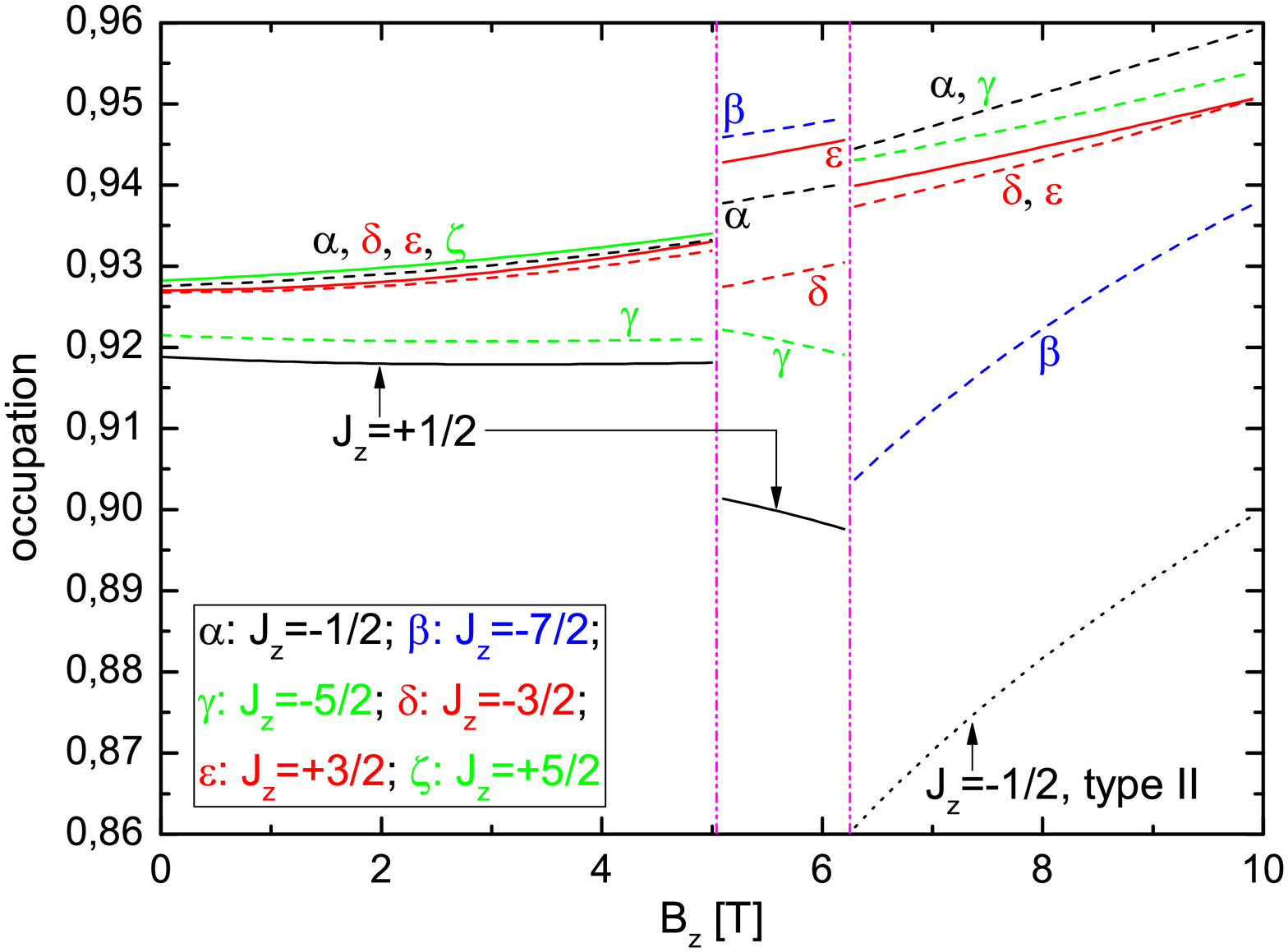}}
\end{tabular}}
\caption{(a-e) The occupation of single-hole orbitals in multi-hole system in external magnetic field $B_z$; weak confinement. The~case~of: (a) two holes, (b) three holes, (c) four holes, (d) five holes, (e) six holes. Only occupied orbitals are shown. Vertical dash-dot-dot lines correspond to $B_z$ of respective level crossings, as marked in Fig.~\ref{WC-multihole-spectra}\label{WC-wektory}}
\end{figure*}

The analysis of the spectra should be supplemented by an examination of the occupation coefficients of orbitals that are lowest-in-energy (at $B_z=0$)   for the multi-hole ground states -- Fig. \ref{WC-wektory}. However, because of the multiple level crossings occur in the one-particle spectrum, we decided to take eight states into consideration: the same six as in the case of strong confinement plus ($J_z=-\frac{7}{2}, m_{-\frac{7}{2}}=1$) and  ($J_z=-\frac{1}{2}, m_{-\frac{1}{2}}=2$) additionally.\cite{annotation3} This would lead to appearance of too many lines if the figures were prepared in the format of Fig.~\ref{wektory}, so only the occupied orbitals are shown.

If two holes are confined in the dot, the state with $J2H_z=0$ is the ground state as shown in Fig.~\ref{WC-multihole-spectra}(a). This state is strongly separated from the excited states, which corresponds to the separation between the first two single-hole levels and next-in-energy ones in Fig.~\ref{WC-onehole}. As expected, the lowest two single-hole orbitals are occupied, which can be seen in Fig.~\ref{WC-wektory}(a). Because of this character of one-particle spectrum, the $N_P=2$ case obviously follows the Aufbau principle.

In the case of three holes, for low magnetic field $J3H_z=-\frac{5}{2}$ state is the ground level of the system. However, at the $B_z=7.4$ T there is a crossing and for larger magnetic fields the $J3H_z=-\frac{5}{2}$ state becomes the ground state. In terms of level occupation [Fig.~\ref{WC-wektory}(b)] the crossing is a transfer of an electron from the $J_z=-\frac{5}{2}$ orbital to the $J_z=-\frac{7}{2}$ one. This corresponds to the $B_6$ crossing -- shown in Fig.~\ref{WC-onehole}(b) and Table~\ref{TabCrossings} -- that takes place between the two single-hole levels in question at a similar magnetic field intensity ($8$ T). It should be noted that there is no trace in $N_P=3$ spectrum and/or orbital occupation of neither the single-hole level crossing at $B_0$ nor the near-degeneracy of relevant levels. Moreover, when no magnetic field is present in the system, the $J3H_z={\pm}\frac{5}{2}$ states have noticeably lower energy than $J3H_z={\pm}\frac{1}{2}$ ones while in single-hole spectrum the $J_z={\pm}\frac{5}{2}$ states have slightly higher energy than $J_z={\pm}\frac{1}{2}$ for $B_z=0$. In both the $J3H_z=-\frac{1}{2}$ and $J3H_z=-\frac{5}{2}$ states the $J_z={\pm}\frac{3}{2}$ orbitals are occupied. Hence, mentioned facts suggest that the $J_z=-\frac{5}{2}$ level is strongly preferred to the $J_z=-\frac{1}{2}$ one in respect to Coulomb repulsion between it and $J_z={\pm}\frac{3}{2}$ orbitals (just like in the case of the case of $N_P=3$ for the system with strong confinement -- see the description in the Sec. \ref{subsubsec:SC-Energy-spectra-of-multiple-holes}). In context of the Aufbau principle we should note: i) the absence of the $B_0$ level crossing clearly marks its violation -- it may be thought of as a "strong" kind of violation, the same kind as described earlier for strong confinement $N_P\in\lbrace3,4\rbrace$ spectra ii) the shift between the magnetic field value of the $J3H_z=-\frac{5}{2}$/$J3H_z=-\frac{7}{2}$ and $J_z=-\frac{5}{2}$/$J_z=-\frac{7}{2}$ crossings which is a more subtle effect -- a "weak" violation of the Aufbau principle.

When one more hole is added to quantum dot [$N_P=4$; Fig.~\ref{WC-multihole-spectra}(c)] then for $B_z{\in}(0,6.49)$ T the $J4H_z=-3$ level is the ground state of the system. Within this interval, the $J_z=\pm\frac{3}{2}$ orbitals are occupied as well as the \textit{lower-energy pair}. At $B_z=6.49$ this level crosses the $J4H_z=-6$ one, which corresponds to a transfer of an electron from the $J_z=-1/2$ orbital to the $J_z=-7/2$ one -- as shown in Fig. \ref{WC-wektory}(c). This intersection in an effect of the crossing of the relevant single-hole levels at $B_5$ (see Fig.~\ref{WC-onehole}(b) and Table~\ref{TabCrossings}). For magnetic field of $B_z=8.48$ T, there is an another crossing -- this time connected to the $B_8$ crossing in Fig.~\ref{WC-onehole}(b) and Table~\ref{TabCrossings} -- as the $J4H_z$ of the ground state switches to $-4$, and an electron is transferred from occupying the $J_z=-\frac{5}{2}$ orbital to the $J_z=-\frac{1}{2}$ (type II) one.

When the dot is charged with five particles [Fig.~\ref{WC-multihole-spectra}(d)], then the ground state in the magnetic field interval $B_z\in\lbrace0,4.25\rbrace$ T has $J5H_z=-\frac{5}{2}$. The occupied orbitals are: $J_z\in\lbrace\pm\frac{3}{2},\pm\frac{1}{2}\rbrace$ and $J_z=-\frac{5}{2}$ [Fig.~\ref{WC-wektory}(d)], which are the five lowest-energy ones for low magnetic field. At the point of crossing ($4.25$ T) the $J_z=1/2$ orbital changes to the $J_z=-7/2$ one and that switches $J5H_z$ value to $-\frac{13}{2}$. The equivalent single-hole crossing takes place at $B_2$ (see Fig.~\ref{WC-onehole}(b) and Table~\ref{TabCrossings}). This $J5H_z$ value of ground state does not change for up to $10$ T, but at $B_z=8.72$ T we observe an another crossing in the five-hole spectrum, and by analysing Fig.~\ref{WC-wektory}(d) one can see that it is connected to the crossing of type I and type II $J_z=-1/2$ single-hole states, as seen in Fig.~\ref{WC-onehole}(b) at $B_7$.

For six holes the $J6H_z=0$ level has lowest energy up to $5.05$ T [Fig.~\ref{WC-multihole-spectra}(e)], with $J_z\in\lbrace\pm\frac{3}{2},\pm\frac{1}{2},\pm\frac{5}{2}\rbrace$ orbitals occupied [Fig.~\ref{WC-wektory}(e)]. At that point $J_z=\frac{5}{2}$ orbital is exchanged for the $J_z=-\frac{7}{2}$ one, which corresponds to $B_1$ crossing in Fig.~\ref{WC-onehole}(b) and lowers the $J6H_z$ of the ground state by six. The next crossing, which takes place for magnetic field of $6.22$ T, marks the transfer of an electron from the $J_z=\frac{1}{2}$ orbital to the $J_z=-\frac{1}{2}$ one of type II -- an analog to $B_4$ in single-hole system -- which sets $J6H_z$ to $-7$.

In the cases of $N_P\in\lbrace4,5,6\rbrace$ the Aufbau principle allows the set of occupied one-particle orbitals to be predicted for each inter-crossing interval along with the order in which the occupied orbitals change in the case of each crossing. However, the actual values of magnetic field intensities at which these changes do occur in multi-hole spectra do not coincide exactly with the $B_z$ values of the relevant single-hole crossings. In short the principle is only "weakly" violated in the meaning as described above for $N_P=3$. The details on the subject of comparison of particular crossings are presented in Table \ref{TabCrossingsCompare}. 

\begin{table}[ht]
\begin{center}
    \begin{tabular}{C{25mm}|C{25mm}|C{25mm}} \hline
    $B_z$ at single-hole crossing & $N_P$ for multi-hole crossing & $B_z$ at multi-hole crossing\\ \hline
	$B_0=3.86$ T & $3$ & does not occur \\ \hline
    $B_1=7.51$ T & $6$ & $5.05$ T  \\ \hline
	$B_2=7.66$ T & $5$ & $4.25$ T \\ \hline
 	$B_3=7.76$ T & not applicable & does not occur \\ \hline
    $B_4=7.90$ T & $6$ & $6.22$ T \\ \hline
    $B_5=7.97$ T & $4$ & $6.49$ T \\ \hline
    $B_6=8.00$ T & $3$ & $7.4$ T \\ \hline
    $B_7=8.17$ T & $5$ & $8.72$ T \\ \hline
    $B_8=8.21$ T & $4$ & $8.48$ T \\ \hline 
    \end{tabular}
\end{center}
\caption{The comparison of the single-hole crosings and corresponding multi-hole ones.}\label{TabCrossingsCompare}\end{table}

The absence of crossing that was corresponding to the $B_3$ one is not a surprise, as the latter one involves the seventh and eighth single-hole levels (in order of increasing energy) and hence the former would appear in the multi-hole spectra of $N_P<7$ only in the case of a "strong" violation of the Aufbau principle.

\subsubsection{Chemical potential}

\begin{figure*}[!ht]
 \makebox[\textwidth][c]{
\begin{tabular}{ll}
(a) & (b)\\
\rotatebox{0}{\epsfxsize=70mm \epsfbox[65 35 705 505] {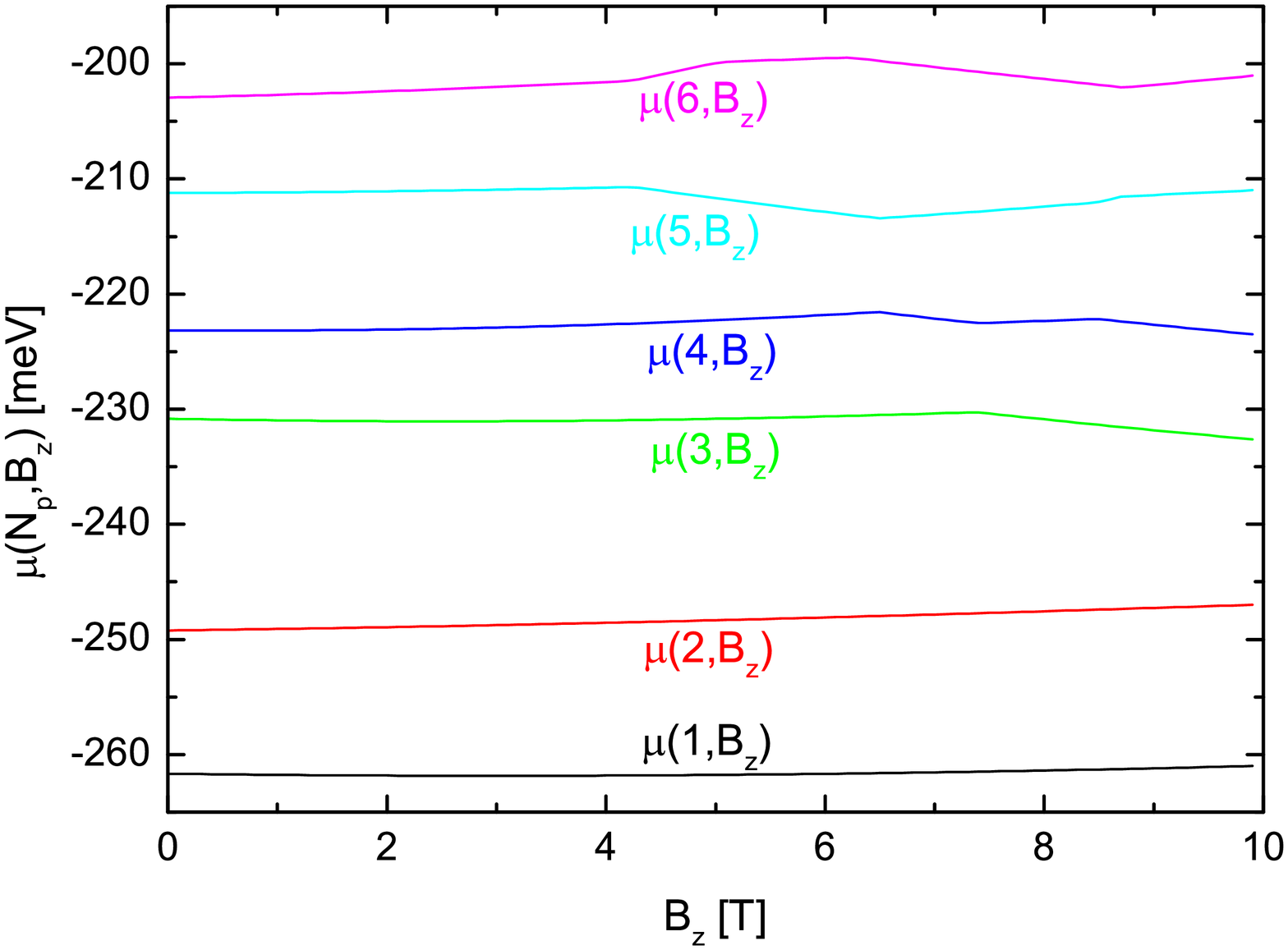}} &
\rotatebox{0}{\epsfxsize=70mm \epsfbox[65 35 705 505] {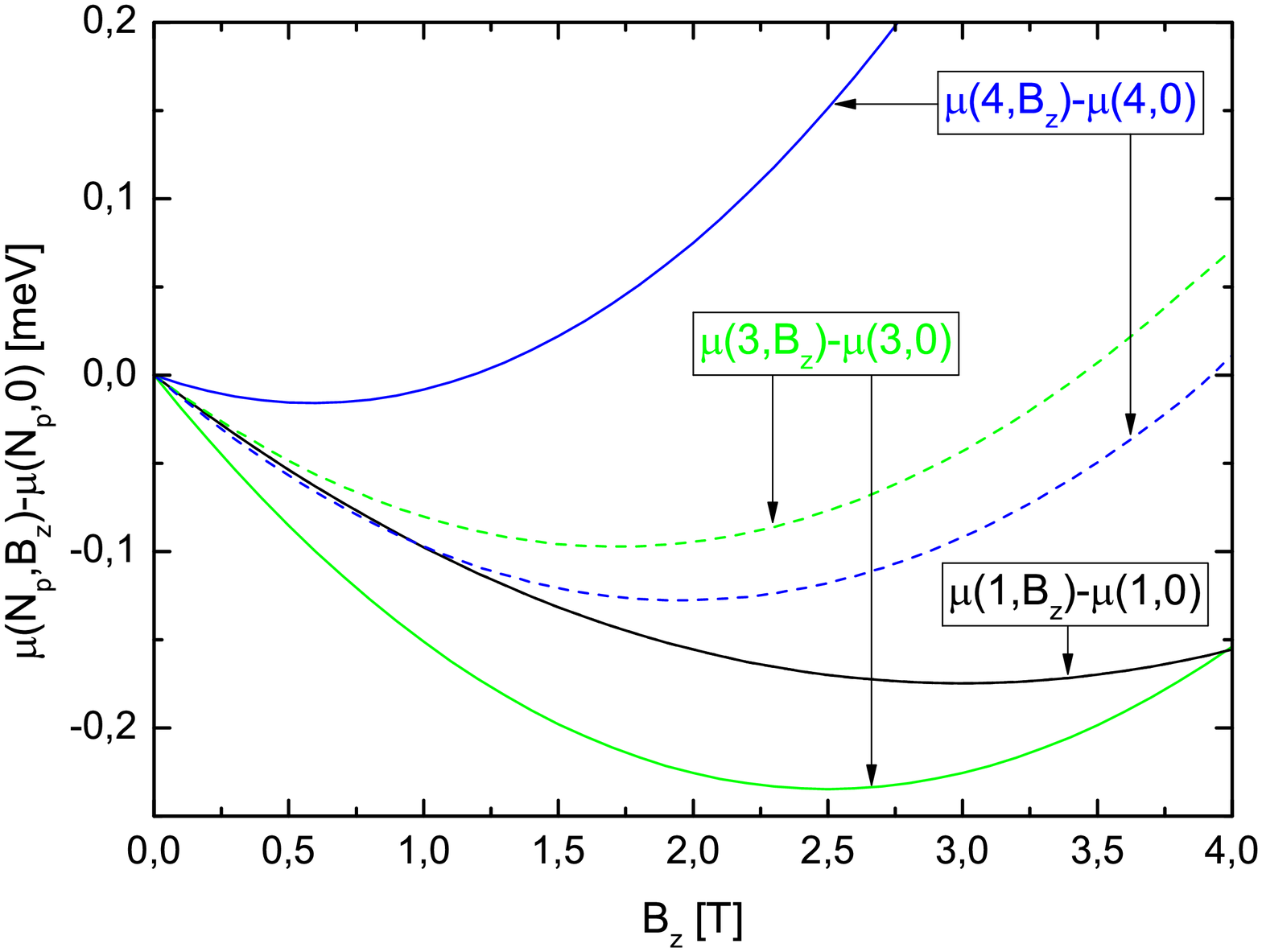}} \\
(c)\\
\rotatebox{0}{\epsfxsize=70mm \epsfbox[65 35 705 505] {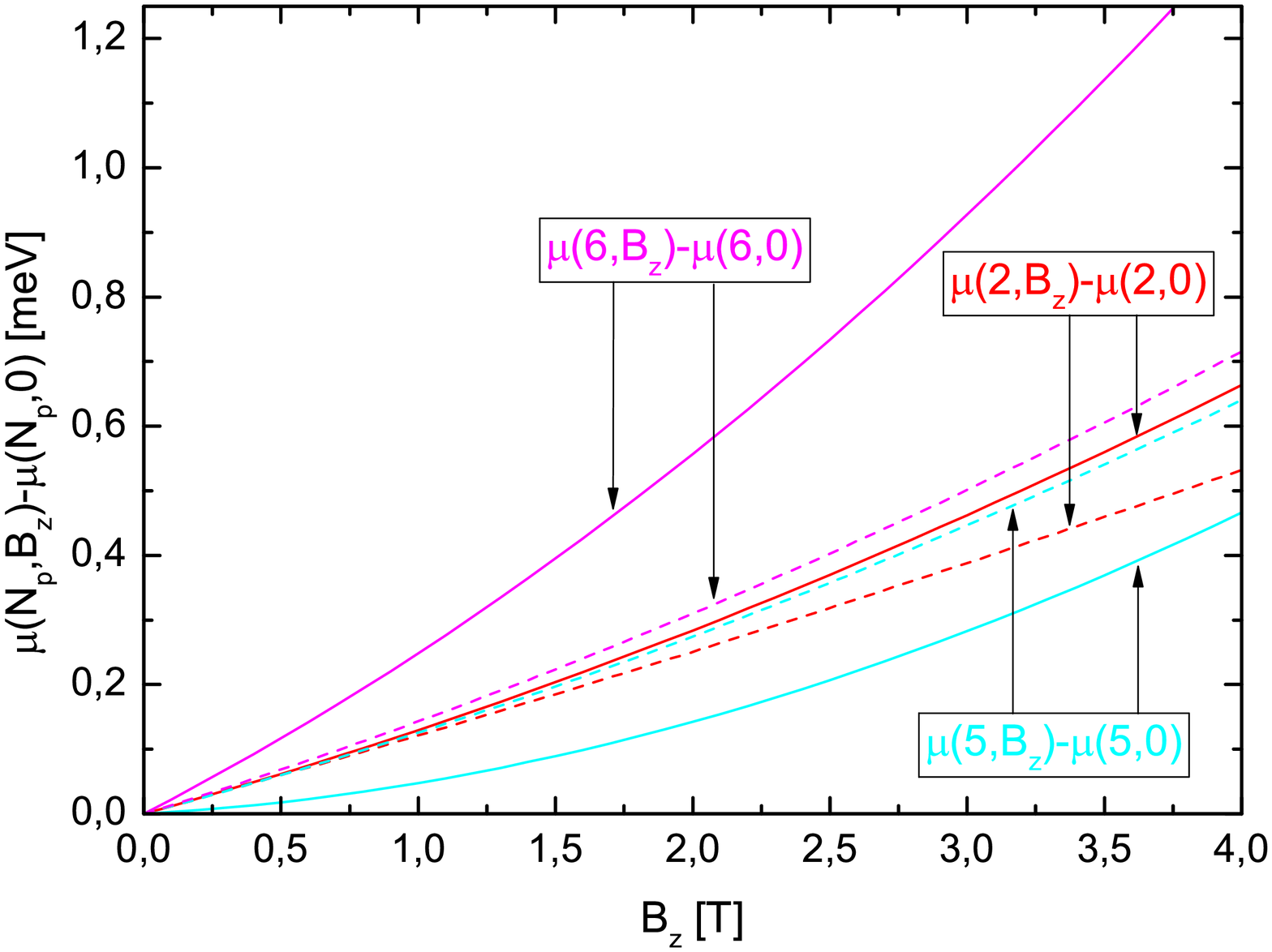}} \\
\end{tabular}}
\caption{(a) The chemical potential as a function of external magnetic field $B_z$ for $N_p{\in}\{1,...,6\}$; weak confinement. \mbox{(b-c)}~Relative dependence of $\mu(N_p,B_z)$ on $B_z$ for: (b) $N_p{\in}\{1,3,4\}$, (c) $N_p{\in}\{2,5,6\}$. Solid lines correspond to the full model, and dashed lines to the non-interacting case.\label{WC-chempot}}
\end{figure*}

The chemical potential ${\mu}(N_p,B_z)=E(N_p,B_z)-E(N_p-1,B_z)$ of the system with the bigger dot is presented in Fig.~\ref{WC-chempot}(a). For $N_P\geq3$ the dependence of ${\mu}(N_p,B_z)$ on external magnetic field in the growth direction shows strong changes for some values of $B_z$. This is an effect of the crossings that take place between the relevant states in multi-hole spectra. As the chemical potential is effectively a difference between two energies, then rapid changes in both $N_P$ and $N_P-1$ spectra are represented in Fig. \ref{WC-chempot}(a). The chemical potential for $N_p=1$ is trivially the same in both models. If one compares the results for \textit{relative} dependence of chemical potential on magnetic field [see Fig.~\ref{chempot}(a,b)] in the case of the normal computation (solid lines) and the non-interacting model (dashed lines), then for $N_p>1$ one can note that $\mu(Np,Bz)-\mu(Np,0)$ have some common characteristics in both models -- unlike for the strong confinement. Specifically, the sign of the values in both models is the same for a sufficiently small magnetic field. In another words -- the signs of the derivatives of $\mu(Np,Bz)-\mu(Np,0)$ at $B_z=0$ agree in both models. The cases for which this derivative is negative are presented in Fig.~\ref{WC-chempot}(b), and the ones with the positive sign in Fig.~\ref{WC-chempot}(c). This is however as far as the similarities go, as the absolute values of the respective derivatives are completely different, with the possible exception of $N_P=2$. For $N_P\in{\lbrace}3,6{\rbrace}$ the value of $\mu(Np,Bz)-\mu(Np,0)$ drifts away from zero much quicker in the full model than in the non-interacting model and for $N_P\in{\lbrace}4,5{\rbrace}$ it drifts away much slower. In the end we reach a similar conclusion as in the system with the strong confinement -- that it is not possible to infer directly the behaviour of the chemical potential in magnetic field $B_z$ without taking the Coulomb interaction in the account.

\section{Discussion and Conclusion}
In Ref. \onlinecite{reuter} an experimental data was presented by Reuter \textit{et al.} for the hole charging spectra of self-assembled InAs quantum dots in perpendicular magnetic fields probed by capacitance-voltage spectroscopy. The authors of that work interpreted the results in the terms of the typical results obtained for electrons \textit{i.e.}~$s$, $p$ and $d$ shell system for envelope functions that conforms to Aufbau principle. They reported so called "incomplete hole shell filling" that means the $d$ shell starts to be occupied before the $p$ shell is full and they understood this as the breaking of the Aufbau principle induced by the Coulomb interaction.

Later, Climente \textit{et al.} presented a work\cite{planelles1} that illustrated the difference between the typical $s$-$p$-$d$ shell system of heavy hole model (similar as in the case of conduction band electrons) and the results obtained when valence band mixing is taken into account. Authors of Ref. \onlinecite{planelles1} have shown that using the latter model for single hole -- in form of the four-band KL Hamiltonian -- leads to obtaining a set of three twofold Kramer degenerate shells. The shells correspond to absolute value of the total angular momentum of the holes $|J_z|$ equal to \mbox{$3/2$, $1/2$} and $5/2$, respectively. In the mentioned work, authors do not include Coulomb interactions in the case of the KL model but instead claim that the behaviour of the system can be understood in terms of non-interacting holes when the valence band mixing is included. The non-interacting picture implies that the Aufbau principle is trivially fulfilled.

This work is a continuation of the studies of the group of Planelles \& Climente [see Refs. \onlinecite{planelles0, planelles1, planelles2, planelles3}]. It employs a model that is in several ways more precise than the one used by Ref. \onlinecite{planelles1}. Firstly spin-orbit split-off bands are included with six-band KL Hamiltonian instead of four-band one. Furthermore, we include the Coulomb interaction between the holes \textit{via} the configuration interaction method. Thirdly, the effect of strain-induced on the confinement potential is taken into account by using the Pikus-Bir Hamiltonian. Finally, an enhanced model for magnetic field is used that is known to avoid some problems of the old model, as described in Sec. \ref{subsec:Magnetic field}.

We show that even if the Aufbau principle would be applied to proper "ladder" of one-particle KL eigenstates (as suggested in Ref. \onlinecite{planelles1}) it still may be violated and that violation is induced by the inter-particle interactions.

Authors of Ref. \onlinecite{reuter} note that  results in case of electron charging experiments could be well explained assuming treating the Coulomb interaction as a perturbation. They also remind that valence band holes have larger effective masses than the conduction band electrons and thus the spatial carrier confinement in hole systems is stronger for them and this results in a stronger hole-hole Coulomb interaction and smaller quantization energies. Together with their results this implies that the inter-particle interaction should be included in a more direct manner. On the other hand the result of Ref. \onlinecite{planelles1} could suggest that the opposite is true. The authors of this latter work use a model that completely omits the Coulomb interaction.

We prove that the Coulomb interaction influences very essentially the general behaviour and properties of the system of few/several holes confined in a quantum dot of the kind considered in the work. Thus, we believe that all further $\vec{k}\cdot\vec{p}$ studies of systems similar to this one should include this interaction directly. Furthermore, we show that the sole introduction of valence band mixing at the one-particle level is insufficient to understand the so called "incomplete hole shell filling" phenomenon.

Please note that works Ref. \onlinecite{atom1,atom2} offer an alternative explanation of the phenomenon that is the topic of this work, from a completely different approach - the atomistic scale one.

\section{Acknowledgements}
This work was  supported by the Polish Ministry of Science and Higher Education, by the EU Human Capital Operation Program, Project No. POKL.04.0101-00-434/08-00 and by National Research Centre according to decision DEC-2012/07/N/ST3/03161. This research was supported in part by PL-Grid Infrastructure.

Author want to thank prof.~J. I. Climente from Universidad Jaume I for insightful discussion of his previous works.


\begin{thebibliography}{99}
\bibitem{annotation0} Unless noted otherwise all formulae are written in atomic units.
\bibitem{annotation1} Note that the sign of $a_v$ and $b_v$ values and/or of $P_{\epsilon},Q_{\epsilon}$ is variously defined in various sources. The orientation of the relative signs should be that - in the case of biaxial strain - the confinement potential is deepened in the case of heavy holes and shallowed for light holes (see Ref. \onlinecite{planelles2, pryor, tadic}).
\bibitem{annotation2} For example let us focus on a case of $J3H_z=-9$. One of the possible $[J_z]$ sequences is $(-5,-4,0)$. As any permutation of the sequence leads to formation of the same subset of basis functions - due to applying antisymmetrization later in the process - only one of all possible permutations is taken into account. Furthermore, there are $[J_z]$ sequences with repeating values, like $(-4,-4,-1)$ or $(-3,-3,-3)$. In those cases swapping $m_{J_z}$ for those equal $J_z$ values gives an equivalent sequence.
\bibitem{annotation3} ($J_z=-\frac{1}{2}, m_{-\frac{1}{2}}=2$) should be actually understood as $J_z=-\frac{1}{2}$ type II.
\end{thebibliography}
\end{document}